\documentstyle[pre,aps,eqsecnum,amssymb,multicol,twoside,rotate,epsfig]{revtex}
\begin{document}
\draft
\tighten

\title{Critical dynamics of a uniaxial and dipolar ferromagnet}

\author{S. Henneberger$^{1,2)}$, E. Frey$^{1,3)}$, P.G. Maier$^{1)}$, F.
  Schwabl$^{1)}$, and G.M.  Kalvius$^{2)}$}

\address{$^{1)}$Institut f\"ur Theoretische Physik and
  $^{2)}$Institut f\"ur Kernphysik und Nukleare Festk\"orperphysik\\
  Physik--Department der Technischen Universit\"at M\"unchen, \\
  James--Franck--Stra\ss e, D--85747 Garching, Germany \\
  $^{3)}$Lyman Laboratory of Physics, Harvard University, Cambridge,
  MA 02138, USA}

\date{\today}
\maketitle

\begin{abstract}
  We study the critical dynamics of three-dimensional ferromagnets
  with uniaxial anisotropy by taking into account exchange and
  dipole-dipole interaction.  The dynamic spin correlation functions
  and the transport coefficients are calculated within a mode coupling
  theory. It is found that the crossover scenario is determined by the
  subtle interplay between three length scales: the correlation
  length, the dipolar and uniaxial wave vector.  We compare our
  theoretical findings with hyperfine interaction experiments on Gd
  and find quantitative agreement. This analysis allows us to identify
  the universality class for Gd. It also turns out that the $\mu$SR
  relaxation rate can be best fitted if it is assumed that muons
  occupy octahedral interstitials sites within the Gd lattice.
\end{abstract}


\begin{multicols}{2}
\narrowtext

\section{Introduction}
\label{introduction}

The spin dynamics of simple ferromagnets in the vicinity to their
Curie point is an archetypical example for critical dynamic phenomena
near second-order phase transitions. In recent years it became
increasingly clear that the dipolar interaction has a dramatic effect on
the critical spin fluctuations of all real ferromagnets
\cite{koetzler:76,mezei:82,koetzler:83,goerlitz:98}. A mode coupling
theory (for a review see e.g. Ref.~\cite{frey-schwabl:94}) including
the dipolar coupling has lead to remarkable success in explaining the
experimental data for materials with cubic lattice structure (e.g. Fe,
Ni, EuO, and EuS).

In this paper we extend this analysis to anisotropic magnetic systems.
A short account of part of these results with emphasis on the
interpretation of experiments on Gd has been given recently
\cite{frey-etal:97}.  Anisotropy acts to suppress critical
fluctuations perpendicular to the easy axis of magnetization and
breaks certain local conservation laws.  This has a marked effect on
the critical dynamics of spin fluctuations near $T_c$. Combining the
effect resulting from magneto-crystalline anisotropy and long-ranged
dipolar interaction one expects a crossover scaling behavior with {\em
  three scaling variables}. Besides the correlation length $\xi$ there
are two more length scales resulting from the strength of the dipolar
interaction and the anisotropy energy with respect to the exchange
energy.

Dipolar interaction is known to be a relevant perturbation with
respect to the fixed point of a $n$-component Heisenberg model, where
the spins are coupled by short-range exchange interaction. It drives
the system to a new dipolar fixed point, whose nature dramatically
depends on the number $n$ of components of the order parameter. For
isotropic Heisenberg ferromagnets ($n=3$) the resulting isotropic
dipolar fixed point is characterized by a set of critical exponents
which are only slightly different from the corresponding values at the
(isotropic) Heisenberg fixed point. The consequences of this crossover
on the static and dynamic correlation functions is by now well known
and has been reviewed recently \cite{frey-schwabl:94}. For uniaxial
($n=1$) systems, however, it was shown by Larkin and Khmelnitskii
\cite{larkin-khmelnitskii:69} that dipolar interaction asymptotically
leads to classical critical behavior with logarithmic corrections in
three dimensions. The asymptotic behavior of this system was also
studied by means of renormalization group theory
\cite{aharony:73,aharony-halperin:75,brezin-zinn-justin:76}, which
revealed that the one loop calculation agrees with the asymptotic
results of Ref.~\cite{larkin-khmelnitskii:69}. The complete crossover
from Ising behavior with non classical exponents to asymptotic
uniaxial dipolar behavior has also recently been analyzed within a
generalized minimal subtraction method \cite{frey-schwabl:90}.

In anisotropic ferromagnets with an easy axis anisotropy there are
two relevant perturbations, the dipolar coupling $g$ and the
anisotropy parameter $m$. As a consequence upon approaching the
critical temperature the system passes through a considerably more
complex crossover region before it reaches its asymptotic critical
behavior.  Four nontrivial fixed points determine the flow of the
various model parameters: the isotropic Heisenberg, the uniaxial
Ising, the isotropic dipolar and the uniaxial dipolar fixed point.
Depending on the relative strength of the dipolar and uniaxial
anisotropy different scenarios are possible. For $g>m$ the system is
supposed to show a crossover cascade from isotropic Heisenberg to
isotropic dipolar to uniaxial dipolar critical behavior. For $m>g$ the
system first crosses over from the Heisenberg to the Ising fixed point
before it turns to the asymptotically stable uniaxial dipolar fixed
point. Both cases seem to be realized in nature; e.g. one finds $m
\approx 1.2 \, 10^{-2}$ and $g \approx 2.0 \, 10^{-3}$ for
Fe$_{14}$Nd$_2$B (see Ref.~\cite{ried-koehler-kronmueller:92}, whereas
$m< g$ in Gd (see below)).

In this paper we study the dynamics of such anisotropic ferromagnets
under the combined influence of dipolar interaction and
magnetocrystalline anisotropy. We proceed as follows. In
Section~\ref{model} we discuss the model Hamiltonian and the effect of
dipolar interaction on the critical behavior in cubic and hexagonal
lattices. We define the dimensionless parameters characterizing the
strength of the dipolar and the uniaxial anisotropy. In
Sec.~\ref{statics} we discuss the static critical behavior of uniaxial
dipolar ferromagnets and derive the eigenvalues and eigenvectors of
the static susceptibility matrix.  The mode coupling theory in terms
of these eigenmodes is formulated in Sec.~\ref{mode_coupling}. We
briefly discuss the limits of an isotropic Heisenberg magnet
without dipolar interaction and of an isotropic Heisenberg magnet
including dipolar interaction. The general case of an uniaxial and
dipolar ferromagnet is discussed in detail.  The mode coupling
equations are solved analytically in certain limiting regions of
parameter space and analytically for intermediate parameter values.
The resulting crossover scenarios for dominating magneto-crystalline
and dipolar anisotropy are discussed, respectively. In
Sec.~\ref{experiment} we compare our theoretical findings with results
from various hyperfine interaction experiments on Gd.  In
Sec.~\ref{summary} we give a summary and discussion of the results.
Some of the technical details of the theory are collected in the
appendix.

\section{The model of an uniaxial dipolar ferromagnet}
\label{model}

We consider a system with $N$ identical spins fixed on the sites of a
three dimensional lattice. Taking into account magneto-crystalline
anisotropy as well as dipolar interaction it is described by a
Heisenberg hamiltonian
\begin{eqnarray}
H = &-&\sum_{i \neq j}
\Biggl\{
\left[ J_{i j}^{\perp} \left( S^x_i S^x_j + S^y_i S^y_j \right) +
J_{i j}^{\parallel} S^z_i S^z_j \right] \nonumber \\
&+& D_{ij}^{\alpha \beta} S_i^\alpha S_j^\beta \Biggr\}.
\label{hamiltonian}
\end{eqnarray}
The magnitude of the magneto-crystalline anisotropy of the system is
given by $\Delta = J^{\parallel} / J^{\perp}$. Here we focus on
uniaxial anisotropy, $\Delta > 1$, with the easy axis of magnetization
along the $z$-axis. The dipolar interaction is characterized by the
tensor
\begin{equation}
D_{ij}^{\alpha \beta} =
- \frac{(g_L \mu_B)^2}{2}
\left( \frac{\delta_{\alpha \beta}}{|{\bf x}_{ij}|^3}
     - \frac{3 x_{ij}^\alpha x_{ij}^\beta}{|{\bf x}_{ij}|^5}
 \right) ,
\end{equation}
with ${\bf x}_{ij} = {\bf x}_i - {\bf x}_j$, $g_L$ the Land\'e factor,
and $\mu_B$ the Bohr magneton.  As shown by Cohen and Keffer
\cite{cohen-keffer:55} dipolar lattice sums
\begin{equation}
 D_{\bf q}^{\alpha \beta} = \sum_{i \neq j}
                            D_{ij}^{\alpha \beta}
                            e^{i {\bf q} \cdot {\bf x}_i }
\end{equation}
can be evaluated by using Ewald's method \cite{born-huang:54}. For
infinite three--dimensional {\em cubic lattices} one finds to leading
order in the wave vector ${\bf q}$ \cite{aharony-fisher:73}
\begin{eqnarray}
 D_{\bf q}^{\alpha \beta}  = \frac{(g_L \mu_B)^2}{2 v_a}
 \Biggl\{
 &&{4 \pi \over 3}
 \left( \delta_{\alpha \beta} - {3 q_\alpha q_\beta \over q^2} \right)
 + \alpha_1 q_\alpha q_\beta \nonumber \\
 + &&\left[\alpha_2 q^2 - \alpha_3 (q_\alpha)^2 \right] \delta_{\alpha \beta}
\Biggr\}  \, ,
\end{eqnarray}
where $v_a$ is the volume of the primitive unit cell with lattice
constant $a$, and $\alpha_i$ are constants, which depend on the
lattice structure (see e.g. \cite{frey-schwabl:94}). Upon expanding
the exchange interaction,
\begin{equation}
J_{\bf q} = {\sum_i}^\prime J_{i0}
            e^{i {\bf q} \cdot {\bf x}_i }
            \approx J_0 - J q^2 a^2 + {\cal O}(q^4) \, ,
\end{equation}
and keeping only those terms, which are relevant in the sense of
renormalization-group theory this results in the following effective
Hamiltonian for dipolar ferromagnets
\begin{eqnarray}
  H = \sum_{\bf q}
      \left[ \left(-J_0 + J q^2 a^2\right) \delta_{\alpha \beta}
             + J g  \frac{q_\alpha q_\beta}{q^2}
      \right]
      S^\alpha_{- \bf q} S^\beta_{\bf q} \, ,
\end{eqnarray}
where the Fourier--transform of the spin variables is defined by
\begin{equation}
 S_{\bf q}^\alpha = {1 \over N} \sum_i S_i^\alpha
 e^{i {\bf q} \cdot {\bf x}_i } \, .
\end{equation}
Here we have defined a dimensionless quantity $g$ as the ratio of the
dipolar energy $(g_L \mu_B)^2/a^3$ and exchange energy $2J$, multiplied by a
factor ${4 \pi a^3 / v_a}$, which depends on the lattice structure.
\begin{equation}
 g = {4 \pi a^3 \over v_a} \, {(g_L \mu_B)^2 / a^3  \over 2 J}
 \propto {{\rm Dipolar \, Energy} \over {\rm Exchange \, Energy}} \, .
\label{dipolar_coupling}
\end{equation}
Strictly speaking there are dipolar corrections of order ${\cal
  O}(q^2)$ to the exchange coupling. But, those can be neglected,
since the strength of the dipolar interaction is small compared with
the exchange interaction.

For Bravais lattices with a {\em hexagonal-closed packed (hcp)
  structure} the dipolar tensor to leading order in ${\bf q}$ becomes
of the form (see Ref.~\cite{fujiki:87} and the appendix)
\begin{eqnarray}
 D_{\bf q}^{\alpha \beta} = \frac{(g_L \mu_B)^2}{2 v_a}
 \Biggl\{
 \!&-& 4 \pi \frac{q_\alpha q_\beta}{q^2}
  + \beta_1^{\alpha \beta} q_\beta q_\beta + \beta_2^\alpha q^2
  \!-\! \beta_3^\alpha (q_\alpha)^2 \nonumber \\
 &+&\beta_4^\alpha
  - \beta_z (q_z)^2 (1 - \delta_{\alpha z})
 \Biggr\}  \, ,
\end{eqnarray}
with the coefficients given in table \ref{coefficients_hcp}. In the
same way as above we find
\begin{eqnarray}
  H = \sum_{\bf q}
      \Biggl[ \biggl(&-&J_0^\alpha - J m^\alpha + J q^2 a^2
              \biggr) \delta_{\alpha \beta} \nonumber \\
                     &+& J g  \frac{q_\alpha q_\beta}{q^2} \Biggr]
      S^\alpha_{- \bf q} S^\beta_{\bf q} \, ,
\label{effective_hamiltonian}
\end{eqnarray}
where the contribution of the dipolar interaction to
the uniaxial anisotropy is given by
\begin{eqnarray}
  m^\alpha = \frac{1}{J} \, \frac{(g_L \mu_B)^2}{2 v_a} \, \beta_4^\alpha .
\end{eqnarray}
Note that the term proportional to $q_\alpha q_\beta / q^2$ depends on
the lattice structure only via the volume of the unit cell. Therefore
the value of $g$ is identical to Eq.~\ref{dipolar_coupling}.

There are two sources of uniaxial anisotropy in the Hamiltonian, Eq.
\ref{effective_hamiltonian}, {\em magnetocrystalline anisotropy}
$J_0^\alpha$, defined by $J^{\alpha}_{{\bf q}}\approx 
J^{\alpha}_{0}-Jq^{2}a^{2}+{\cal O}(q^{4})$, and $m^\alpha$ which characterizes the 
{\em dipole-dipole interaction}.  {\em Crystal
  field} contributions are expected to be the dominant factor in
systems like LiTbF$_4$ \cite{holmes:75} and Fe$_{14}$Nd$_2$B
\cite{ried-koehler-kronmueller:92}.

In addition, the dipolar interaction introduces an anisotropy of the
spin-fluctuations with respect to the wave vector ${\bf q}$ which is
reflected by the term proportional to $q_\alpha q_\beta / q^2$.  The
magnitude $g$ of this anisotropy is given by $g = 4 \pi (g_L \mu_B)^2
/ 2 J v_a$.  We define a dimensionless quantity
\begin{eqnarray}
m = (g_L \mu_B)^2 (\beta_4^\parallel - \beta_4^\perp) / 2 J v_a
\end{eqnarray}
proportional to the ratio between the anisotropy energy and the
exchange energy.  Putting in values for Gd the ratio of the dipolar
contribution to the term $q_\alpha q_\beta / q^2$ and to the uniaxial
anisotropy is $\sqrt{g/m} = 7.8738$.  In section \ref{experiment} we
will show that all available data for Gd can be explained by assuming
that the uniaxial anisotropy is solely due to the dipolar interaction.

\section{The critical static behavior}
\label{statics}

\subsection{The static susceptibility}

Upon using standard techniques such as the Hubbard-Stra\-tono\-vich
transformation one may derive an effective Landau-Ginzburg free
energy functional from the microscopic Hamiltonian,
Eq.~(\ref{hamiltonian}).  Then the renormalized free energy in
Gaussian approximation becomes,
\begin{equation}
H =  \frac{1}{2} \int_{{\bf k}} \chi^{-1}_{\alpha \beta}({\bf k})
\, S^{\alpha}({\bf k})S^{\beta}(-{\bf k}),
\end{equation}
where $\int_{\bf k} = v_a \int d^3 k / (2 \pi)^3$. The inverse
renormalized propagator (susceptibility) is given by
\begin{equation}
\chi^{-1}_{\alpha \beta}({\bf k}) = J \left[ (r_{\alpha} +a^{2}k^2)
\, \delta_{\alpha \beta} +
g \, \frac{k_{\alpha} k_{\beta}}{k^2} \right],
\label{propagator}
\end{equation}
where we have taken an Ornstein-Zernike functional form and assumed
that the static crossover can be described in terms of scale dependent
parameteres $r_\alpha$ and $g$. In other words, all the crossover is
contained in effective static exponents. We have chosen the reference
frame such that the $z$-axis coincides with the easy axis of
magnetization. Hence the ``mass'' of the corresponding spin
fluctuations,
\begin{eqnarray}
r_z & = & r \, = \, \frac{a^{2}}{\xi^2},
\label{easy_parameter}
\end{eqnarray}
defines the correlation length $\xi = \xi_0 [(T-T_c)/T_c]^{-\nu}$ with
a non-universal amplitude $\xi_0$ and the effective exponent $\nu$. In
the hard sector the ``masses'' do not vanish upon approaching the
critical temperature, but saturate at a finite value $m$
\begin{eqnarray}
r_x & = & r_y \, =  \, r + m  \, = \, \frac{a^{2}}{\xi^2} + a^{2} q_{_A}^2.
\label{hard_parameter}
\end{eqnarray}
The parameter $m=(q_{_A} a)^2$ characterizes
the magnetocrystalline anisotropy.  The ratio of dipolar to exchange
interaction can be described by the Parameter $g = (q_{_D}a)^2$. 
Thus, in addition to the correlation length there are two more
relevant length scales, $q_{_A}^{-1}$ and $q_{_D}^{-1}$.  With the
scaling variables, ${\bf R} = (x,y,z)$,
\begin{eqnarray}
x = \frac{1}{q \xi}, \quad y = \frac{q_{_D}}{q}, \quad z = \frac{q_{_A}}{q}
\end{eqnarray}
the susceptibility matrix becomes,
\begin{eqnarray}
\chi_{\alpha \beta}({\bf k})
 =  \chi_{\alpha \beta}({\bf k}; \xi, q_{_D}, q_{_A})
 = \frac{1}{J} \, (aq)^{-2} \, \hat{\chi}_{\alpha \beta}({\hat {\bf k}};{\bf R}),
\end{eqnarray}
with
\begin{eqnarray}
{\hat \chi}^{-1}_{\alpha \beta} &=& \left(
\begin{array}{ccc} \scriptstyle
 1 +  x^2  +  z^2  &  \scriptstyle 0 &  \scriptstyle 0 \\
 \scriptstyle 0 &  \scriptstyle 1 + x^2 + z^2  &  \scriptstyle 0 \\
 \scriptstyle  0 &  \scriptstyle 0 &  \scriptstyle 1 + x^2
\end{array}
\right)_{\alpha \beta}  +  y^2 \frac{q_\alpha q_\beta}{q^2}
\end{eqnarray}
For a more complete discussion of the static crossover and a
calculation of the effective exponents we refer the reader to the
literature \cite{frey-schwabl:90,frey-schwabl:91,%
  ried-millev-faehnle-kronmueller:96a,ried-millev-faehnle-kronmueller:96b}.

\subsection{Eigenvalues and eigenvectors}
To investigate the physical properties of the static susceptibility
one has to find the eigenvectors and eigenvalues of the spin system.
>From Eq.~(\ref{propagator})--(\ref{hard_parameter}) the eigenvalues of
the inverse susceptibility matrix are found to be
\begin{eqnarray}
\lambda_1({\bf k}) & = & k^2 + \xi^{-2} + q_{_A}^2
\label{eigen1} \\
\lambda_2({\bf k}) & = &k^2 + \xi^{-2} +
\frac{1}{2} \left[q_{_D}^2 + q_{_A}^2 + W \right]
\label{eigen2} \\
\lambda_3({\bf k}) & = &k^2 + \xi^{-2} +
\frac{1}{2} \left[q_{_D}^2 + q_{_A}^2 - W \right]
\label{eigen3}
\end{eqnarray}
where $W = \sqrt{(q_{_D}^2+q_{_A}^2)^2-4 q_{_D}^2 q_{_A}^2 k_3^2/k^2}$.
The corresponding eigenvectors are given by
\begin{eqnarray}
{\bf e}_1({\bf k}) & = &
\mbox{\large $ \left(
\begin{array}{c}
\frac{k_2}{\sqrt{k_1^2 + k_2^2}} \\
-\frac{k_1}{\sqrt{k_1^2 + k_2^2}} \\
0
\end{array} 
\right) $ }  \mbox{\small ${\rm sgn} \left( k_1 \right)  $ }, \\
{\bf e}_2({\bf k}) & = &
\mbox{\large $ \left( 
\begin{array}{c}
\frac{k_1}{\sqrt{k_1^2 + k_2^2 + f_{+}^2({\bf k})}}  \\
\frac{k_2}{\sqrt{k_1^2 + k_2^2 + f_{+}^2({\bf k})}}  \\
\frac{f_{+}({\bf k})}{\sqrt{k_1^2 + k_2^2 +
f_{+}^2({\bf k})}}  \end{array}
\right) $ }  \mbox{\small ${\rm sgn} \left( f_{+}({\bf k}) \right)
 $ } , \\
{\bf e}_3({\bf k}) & = &
\mbox{\large $ \left( 
\begin{array}{c}
\frac{k_1}{\sqrt{k_1^2 + k_2^2 + f_{-}^2({\bf k})}} \\
\frac{k_2}{\sqrt{k_1^2 + k_2^2 + f_{-}^2({\bf k})}}  \\
\frac{f_{-}({\bf k})}{\sqrt{k_1^2 + k_2^2 +
f_{-}^2({\bf k})}}  \end{array}
\right) $ }  \mbox{\small ${\rm sgn} \left( f_{-}({\bf k})
\right)  $ } ,
\end{eqnarray}
with
\begin{equation}
f_{\pm}({\bf k}) =
 \frac{2 k_3 (1 - k_3^2 / k^2) q_{_D}^2}
      {q_{_A}^2 + (1 - 2 k_3^2/ k^2 ) q_{_D}^2
       \pm W} .
\end{equation}
This eigenvectors obey the symmetry relation $ {\bf e}_{\alpha}({\bf
  k}) = {\bf e}_{\alpha}(-{\bf k})$.  It is interesting to note that
due to the combined effect of the dipolar interaction and the uniaxial
anisotropy the eigenvalues of the susceptibility matrix remain finite
in the limit $k \rightarrow 0$ and upon approaching the critical
temperature. Only if the angle $\nu$ between the easy axis of
magnetization and the wave vector is $\nu = 90^o$ the third eigenvalue
becomes critical. In the latter case the eigenvectors reduce to ${\bf
  e}_1 ({\bf k}) = (k_2,k_1,0)$, ${\bf e}_2 ({\bf k}) = (k_1,k_2,0)$
and ${\bf e}_3 ({\bf k}) = (0,0,1)$, i.e. the critical eigenvector is
along the easy axis of magnetization.

\begin{figure}[htb]
  \centerline{\epsfxsize=0.5\columnwidth
    \rotate{\rotate{\rotate{\epsfbox{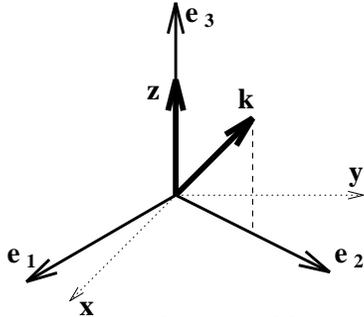}}}}}
\caption{Eigenvectors in the uniaxial limit. The easy axis of magnetization
  points along the z-axis. The eigendirections in the uniaxial case
  are the easy axis of magnetization and two directions perpendicular
  to the easy axis, which here are chosen such that one is the
  projection of the wave vector into the xy-plane.}
\label{figure1}
\end{figure}

In order to understand the critical behavior of the static
susceptibilities it is useful to consider the uniaxial and the dipolar
limit. If the dipolar anisotropy can be neglected with respect to the
uniaxial anisotropy, $q_{_D} \ll q_{_A}$, the eigenvectors reduce to
(see figure \ref{figure1})
\begin{eqnarray}
{\bf e}_1({\bf k})  &=&
\mbox{\large $ \left( 
\begin{array}{c}
\frac{k_2}{\sqrt{k_1^2 + k_2^2}}  \\
\frac{-k_1}{\sqrt{k_1^2 + k_2^2}}  \\
0  \end{array} 
\right) \mbox{\small ${\rm sgn}(k_1)$ }$} , \\
{\bf e}_2({\bf k})  &=&
\mbox{\large $ \left( 
\begin{array}{c}
\frac{k_1}{\sqrt{k_1^2 + k_2^2}}  \\
\frac{k_2}{\sqrt{k_1^2 + k_2^2}}  \\
0  \end{array} 
\right) \mbox{\small ${\rm sgn}(k_1)$ }$} , \\
{\bf e}_3({\bf k})  &=&
\left( 
\begin{array}{c}
0 \\
0 \\
1 \end{array} 
\right) .
\end{eqnarray}

The corresponding eigenvalues are given by
\begin{eqnarray}
\lambda_1({k}) & = & k^2 + \xi^{-2} + q_{_A}^2, \\
\lambda_2({k}) & = & k^2 + \xi^{-2} + q_{_A}^2, \\
\lambda_3({k}) & = & k^2 + \xi^{-2}.
\end{eqnarray}
In the pure dipolar limit, $q_{_A} = 0$, one finds the eigenvectors
\begin{eqnarray}
{\bf e}_1({\bf k})  &=&
\mbox{\large $ \left( 
\begin{array}{c}
\frac{k_2}{\sqrt{k_1^2 + k_2^2}}  \\
\frac{-k_1}{\sqrt{k_1^2 + k_2^2}}  \\
0  \end{array} 
\right) {\rm sgn}(k_1) $} , \\
{\bf e}_2({\bf k}) & = & \frac{{\bf k}}{k} \, {\rm sgn}(k_3) , \\
{\bf e}_3({\bf k}) & = &
\mbox{\large $ \left( 
\begin{array}{c}
- \frac{k_1 \, k_3}{k \sqrt{k_1^2 + k_2^2}}  \\
- \frac{k_2 \, k_3}{k \sqrt{k_1^2 + k_2^2}}  \\
\frac{k_1^2 + k_2^2}{k \sqrt{k_1^2 + k_2^2}}
 \end{array} 
\right) $} ,
\end{eqnarray}
where ${\bf e}_2 ({\bf k})$ points along the wave vector ${\bf k}$ and ${\bf
e}_1 ({\bf k})$ is perpendicaular to the wave vector ${\bf k}$ and the easy
axis of magnetization. The third eigenvector is perpendicular to ${\bf k}$ and
${\bf e}_1$.

\begin{figure}[htb]
  \centerline{\epsfxsize=0.5\columnwidth
    \rotate{\rotate{\rotate{\epsfbox{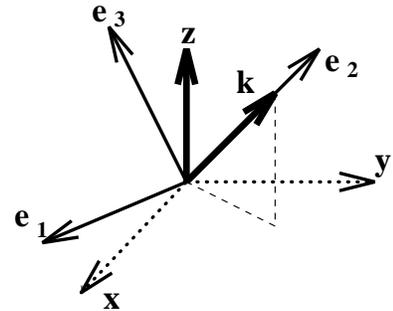}}}}}
\caption{Eigenvectors in the dipolar limit. The easy axis of magnetization
  points along the z-axis. The eigendirections in the dipolar case are
  paralell and perpendicular to the wave vector ${\bf k}$.}
\label{figure2}
\end{figure}

The corresponding eigenvalues
\begin{eqnarray}
\lambda_1({\bf k})  & = &
k^2 + \xi^{-2}  \quad \hspace{9mm}  \mbox{(transverse $T_1$),} \\
\lambda_2({\bf k}) & = &
k^2 + \xi^{-2} + q_{_D}^2 \quad \mbox{(longitudinal $L$),} \\
\lambda_3({\bf k}) & = &
k^2 + \xi^{-2} \quad \hspace{9mm} \mbox{(transverse $T_2$).}
\end{eqnarray}
show that the spin fluctuations transverse to the wave vector ${\bf
  k}$ are critical, but the longitudinal ones stay finite. The dipolar
wave vector $q_{_D}$ determines the crossover scale where the
longitudinal susceptibility turns finite.

\section{Mode coupling theory for the critical dynamics}
\label{mode_coupling}

In this section we calculate the wave vector and frequency dependence
of the spin correlation functions, quantities which are (in principle)
directly accessible to neutron scattering experiments. For this
purpose we use standard mode coupling theory. The basic idea
underlying MC theory is that near the critical point the relevant
dynamics are described through slowly varying macroscopic modes;
i.e.: the conserved quantities and the order parameter. The dynamics
are formulated most conveniently in terms of the Kubo relaxation functions,
\begin{eqnarray}
\Phi_{\alpha \beta} ({\bf q},t) =
  i \lim_{ \epsilon \to 0} \int \limits_{t}^{\infty} d\tau
  e^{- \epsilon \tau} \langle \lbrack s^\alpha ({\bf q}, \tau),
  s^\beta ({\bf q},0)^{\dagger} \rbrack \rangle,
\end{eqnarray}
for the components $s^\alpha ({\bf q},t)$ of the eigenvectors of the
static susceptibility matrix. Here we use the normalization
$\Phi_{\alpha \beta} ({\bf q},t=0) = 1$, i.e., the spin variables are
normalized with respect to the static susceptibilities. Here $\langle
\, \, \, \rangle$ denotes the thermal average and $\lbrack \, , \,
\rbrack$ the commutator. The corresponding frequency dependent
relaxation functions are defined by a half sided Fourier transform
\begin{eqnarray}
\Phi_{\alpha \beta} ({\bf q},\omega) =
  \int \limits_{0}^{\infty} dt e^{i \omega t}
  \Phi_{\alpha \beta} ({\bf q},t).
\end{eqnarray}
This Kubo relaxation function is related to the transport coefficients
$\Gamma_{\alpha \beta} ({\bf q},t)$ and the frequency matrix
$\omega_{\alpha \beta}({\bf q})$ by
\begin{eqnarray}
\frac{d}{dt} \, \Phi_{\alpha \beta} ({\bf q},t) &=&
  i \omega_{\alpha \mu} ({\bf q}) \Phi_{\mu \beta} ({\bf q},t) \nonumber \\
  &\quad& - \int_0^t d \tau \, \Gamma_{\alpha \mu} ({\bf q},t-\tau)
                  \Phi_{\mu \beta} ({\bf q},\tau) \, ,
\end{eqnarray}
where the frequency matrix is given by
\begin{eqnarray}
\omega_{\alpha \beta} ({\bf q}) =
  \frac{1}{\sqrt{\chi_\alpha ({\bf q}) \chi_\beta ({\bf q})}}
  \langle \lbrack s^\alpha ({\bf q}),
                  s^\beta (- {\bf q})
          \rbrack
  \rangle.
\end{eqnarray}
The coefficients, $\Gamma_{\alpha \beta}({\bf q},t)$, of the memory
matrix are determined self-consistently from decay processes of the
spin-modes. If only two mode decay processes are taken into account,
the spin relaxation functions enter quadratically into the coupled
integro-differential equations for the $\Gamma_{\alpha \beta}({\bf
  q},t)$. Frequently one introduces in addition a Lorentzian
approximation for the Kubo relaxation functions, which results in a
simplified set of mode coupling equations for the line widths. For
instance the Resibois-Piette scaling function for isotropic
ferromagnets is obtained on this level of approximation
\cite{resibois-piette:70}.

\subsection{Equations of motion for the eigenmodes}

The analysis of the static susceptibility suggest to decompose the
spin operator in three components along the eigenvectors of the spin
Hamiltonian,
\begin{equation}
{\bf S}_{\bf k} = s^1_{\bf k} {\bf e}_1({\bf k}) +
                  s^2_{\bf k} {\bf e}_2({\bf k}) +
                  s^3_{\bf k} {\bf e}_3({\bf k}) .
\end{equation}
Written out in its cartesian components (which from now on are
indicated by Latin labels) this reads
\begin{eqnarray}
S^i_{\bf k} &=& \sum_\alpha s^\alpha_{\bf k} e_{\alpha i}({\bf k}).
\label{transformation}
\end{eqnarray}
The back transform reads
\begin{eqnarray}
s^{\alpha}_{\bf k} &=&  \sum_i S^i_{\bf k} e_{\alpha i}({\bf k}),
\end{eqnarray}
where the components of the spin operators in the eigenvector basis
are labeled by Greek indices.  With the commutation relations for the
cartesian components of the spin operator one is led to the
commutators of the spin components $s^{\alpha}$,
\begin{eqnarray}
\left[ s^{\alpha}_{-{\bf k}}, s^{\beta}_{\bf q} \right] & = &
- i \hbar \sum_{\gamma}
  U_{\alpha \beta}^\gamma ({\bf k}, {\bf q}) s^\gamma_{{\bf q}\!-\!{\bf k}}
\end{eqnarray}
with
\begin{eqnarray}
U_{\alpha \beta}^\gamma({\bf k}, {\bf q})
= \sum_{ijk} \varepsilon_{ijk} \,
  e_{\alpha i} ({\bf k}) \,
  e_{\beta j} ({\bf q})  \,
  e_{\gamma k} ({\bf q}-{\bf k}),
\end{eqnarray}
where $ \varepsilon_{ijk} $ is the Levi-Cevita symbol. The Heisenberg
equations of motion become
\begin{eqnarray}
\frac{d s^{\alpha}_{\bf q}}{d t} & = & {\dot s}^{\alpha}_{\bf q}
= \frac{i}{\hbar}  \left[ H, s^{\alpha}_{\bf q} \right] \nonumber \\
 & = & \frac{i J }{\hbar} \int_{{\bf k}}  \sum_\beta
\lambda_{\beta}({\bf k}) \left\{ s^{\beta}_{\bf k},
\left[s^{\beta}_{-{\bf k}}, s^{\alpha}_{\bf q} \right] \right\} \nonumber \\
& = &  J \int_{{\bf k}} \sum_{\beta \gamma} \,
\lambda_{\beta}({\bf k}) U_{\beta \alpha}^{\gamma}({\bf k}, {\bf q})
\left\{ s^{\beta}_{\bf k}, s^{\gamma}_{{\bf q}-{\bf k}} \right\} .
\end{eqnarray}
In the classical limit this can be rewritten as
\begin{equation}
\frac{d s^{\alpha}_{\bf q}}{dt} =  2  \, J  \int_{{\bf k}}
 \sum_{\beta \gamma}
T^{\beta \gamma}_{\alpha} ({\bf k}, {\bf q})
s^{\beta}_{\bf k} s^{\gamma}_{{\bf q}-{\bf k}} ,
\end{equation}
with
\begin{eqnarray}
T^{1 1}_{\alpha} ({\bf k}, {\bf q}) & = & \lambda_{1}
({\bf k}) U_{1 \alpha}^{1}({\bf k}, {\bf q}) , \nonumber \\
T^{1 2}_{\alpha} ({\bf k}, {\bf q}) & = & \left( \lambda_{1}
({\bf k})- \lambda_{2}({\bf q}\!-\!{\bf k}) \right)
 U_{1 \alpha}^{2}({\bf k}, {\bf q}) , \nonumber \\
T^{1 3}_{\alpha} ({\bf k}, {\bf q}) & = & \left( \lambda_{1}
({\bf k})- \lambda_{3}({\bf q}\!-\!{\bf k}) \right)
 U_{1 \alpha}^{3}({\bf k}, {\bf q}) , \nonumber \\
T^{2 1}_{\alpha} ({\bf k}, {\bf q}) & = & 0 , \nonumber \\
T^{2 2}_{\alpha} ({\bf k}, {\bf q}) & = & \lambda_{2}
({\bf k}) U_{2 \alpha}^{2}({\bf k}, {\bf q}) , \nonumber \\
T^{2 3}_{\alpha} ({\bf k}, {\bf q}) & = & \left( \lambda_{2}
({\bf k})- \lambda_{3}({\bf q}\!-\!{\bf k}) \right)
U_{2 \alpha}^{3}({\bf k}, {\bf q}) , \nonumber \\
T^{3 1}_{\alpha} ({\bf k}, {\bf q}) & = & 0 , \nonumber \\
T^{3 2}_{\alpha} ({\bf k}, {\bf q}) & = & 0 , \nonumber \\
T^{3 3}_{\alpha} ({\bf k}, {\bf q}) & = & \lambda_{3}
({\bf k}) U_{3 \alpha}^{3}({\bf k}, {\bf q}) ,
\end{eqnarray}
where we have used
\begin{equation}
U_{\alpha \beta}^{\gamma}({\bf k},{\bf q}) =
- U_{\gamma \beta}^{\alpha}({\bf q}\!-\!{\bf k},{\bf q}).
\end{equation}

\subsection{Mode coupling equations}

Using the standard mode coupling formalism for the spin variables
$s^\alpha ({\bf q})$ one gets the following equations for the diagonal
elements of the memory matrix
\begin{eqnarray}
\Gamma_\alpha ({\bf q},t) & = &
\frac{1}{\chi_{\alpha} ({\bf q})}
\left( \dot{s}^{\alpha}_{\bf q}(t),\dot{s}^{\alpha}_{\bf q}(0) \right)
\nonumber \\
& = & \frac{4 \, J^2}{\chi_{\alpha}({\bf q})} \int_{{\bf k}} \int_{{\bf k}'}
T^{\beta \gamma}_{\alpha} ({\bf k}, {\bf q})
\sum_{\beta \gamma} \sum_{\beta' \gamma'}
 \, T^{\beta' \gamma'}_{\alpha} ({\bf k}', {\bf q}) \nonumber \\
 & & \times \left( s^{\beta}_{\bf k} (t)
                   s^{\gamma}_{{\bf q}-{\bf k}} (t),
                   s^{\beta'}_{{\bf k}'} (0)
                   s^{\gamma'}_{{\bf q}-{\bf k}'} (0) \right) .
\end{eqnarray}
The off-diagonal elements are zero since the corresponding relaxation
functions vanish due to the symmetry properties of the Hamiltonian. If
the four-point correlation function on the right hand side is
factorized into a product of two two-point correlation functions, one
gets
\begin{eqnarray}
\Gamma_\alpha ({\bf q},t) & = & \frac{4 k_B T J^2 }
{\chi_{\alpha}({\bf q})}
\int_{{\bf k}} \sum_{\beta \gamma}
K^{\beta \gamma}_{\alpha} ({\bf k}, {\bf q}) \nonumber \\
&& \times \Phi_\beta ({\bf k}, t) \Phi_\gamma ({\bf q}\!-\!{\bf k}, t) ,
\end{eqnarray}
with the vertex functions $ K^{\beta \gamma}_{\alpha} ({\bf k}, {\bf
  q}) $ for the decay of the mode $\alpha$ into the modes $\beta$ and
$\gamma$ given by
\begin{eqnarray}
K^{\beta \beta }_{\alpha} ({\bf k}, {\bf q}) & = &
T^{\beta \beta}_{\alpha} ({\bf k}, {\bf q})
U_{\alpha \beta}^{\beta}({\bf k}, {\bf q})
\left( \lambda_{\beta} ({\bf k}) \! - \!
       \lambda_{\beta}({\bf q}\!-\!{\bf k}) \right) , \nonumber \\
K^{\beta \gamma }_{\alpha} ({\bf k}, {\bf q}) & = &
T^{\beta \gamma}_{\alpha} ({\bf k}, {\bf q}) \,
T^{\beta \gamma}_{\alpha} ({\bf k}, {\bf q}) ,  \quad \beta \not = \gamma,
\end{eqnarray}
The corresponding equation for the Fourier transform reads
\begin{eqnarray}
\Gamma_\alpha ({\bf q},\omega) & = &
\int_{-\infty}^{\infty} dt  \, e^{i \omega t} \,
\Gamma_\alpha ({\bf q}, t) dt \nonumber \\
& = & \frac{4 k_B T J^2}{\chi_{\alpha}({\bf q})}
\int_{{\bf k},\omega'} \sum_{\beta \gamma}
K^{\beta \gamma }_{\alpha} ({\bf k}, {\bf q}) \nonumber \\
&\times& \Phi_\beta ({\bf k}, \omega)
\Phi_\gamma ({\bf q}\!-\!{\bf k}, \omega\! - \!\omega') ,
\end{eqnarray}
where $\int_\omega = \int d \omega / 2 \pi$ and
\begin{equation}
\Phi_\alpha ({\bf q}, \omega) =
\frac{i \chi_{\alpha} ({\bf q})}
     {\omega + i \Gamma_\alpha ({\bf q}, \omega)}
\end{equation}
denotes the half-sided Fourier-transform of $\Phi_\alpha ({\bf q},
t)$.  If the transport coefficients vary only slowly with $\omega$ one
may replace the relaxation functions by simple Lorentzians
\begin{eqnarray}
\Phi_{\alpha}   ({\bf q}, \omega)
=  \frac{i \chi_{\alpha} ({\bf q})}
        {\omega + i \Gamma_{\alpha} ({\bf q})};
\end{eqnarray}
i.e., the transport coefficients are replaced by their values at
$\omega = 0$:
\begin{eqnarray}
\Gamma_{\alpha}({\bf q}) =
\Gamma_{\alpha}({\bf q}, \omega = 0) .
\end{eqnarray}
This additional approximation finally leads to a simplified set of
coupled integral equations for the diagonal elements of the transport
coefficients at zero frequency:
\begin{eqnarray}
\Gamma_\alpha ({\bf q}) &=&
4 k_B T J^2 \int_{{\bf k}}
K^{\beta \gamma }_{\alpha} ({\bf k}, {\bf q})
 \frac{\chi_{\beta}({\bf k}) \chi_{\gamma}({\bf q}\!-\!{\bf k})}
{\chi_{\alpha}({\bf q})} \nonumber \\
&& \times \frac{1}{\Gamma_\beta ({\bf k}) +
\Gamma_\gamma ({\bf q}\!-\!{\bf k})} .
\end{eqnarray}
As emphasized before, the uniaxial anisotropy and the dipolar
interaction introduce two extra length scales $q_{_A}^{-1}$ and
$q_{_D}^{-1}$, respectively.  This entails the following extension of
the static scaling law for the spin susceptibilty (in the eigenvector
basis)
\begin{eqnarray}
  \chi_\alpha ( {\bf q}; \xi, q_{_D}, q_{_A} ) =
  \frac{1}{J q^2} \,
  {\hat \chi}_\alpha ( {\hat {\bf q}}; {\bf R} ),
\label{static_scaling}
\end{eqnarray}
where ${\hat {\bf q}}$ is a unit vector and ${\bf R}=(x,y,z)$ a vector
of scaling variables
\begin{eqnarray}
  x = \frac{1}{q \xi}, \qquad y = \frac{q_{_D}}{q}, \qquad
  {\rm and} \qquad z = \frac{q_{_A}}{q}.
\end{eqnarray}
The suceptibilites are given in terms of the eigenvalues in
Eqs.~(\ref{eigen1})--(\ref{eigen3}) as
\begin{eqnarray}
   {\hat \chi}_\alpha ( {\hat {\bf q}}; {\bf R} ) =
   q^2 \lambda_\alpha^{-1} ({\bf q}; \xi,q_{_D}, q_{_A}).
\end{eqnarray}
Note that the susceptibilities of the spin variables $s^\alpha_{\bf
  q}$ depend not only on the three scaling variables $x$, $y$ and $z$,
but also on the direction ${\hat {\bf q}} = {\bf q} / q$ of the wave
vector ${\bf q}$ with respect to the easy axis of magnetization. Thus,
all scaling functions depend on {\em four variables}. The mode
coupling equations are consistent with the same type of scaling law
for the line widths $\Gamma_\alpha ({\bf q};\xi,q_{_D},q_{_A})$
\begin{equation}
\Gamma_\alpha ({\bf q};\xi,q_{_D},q_{_A}) =
A\, q^{{z}} \, \gamma_{\alpha}({\hat{\bf q}},{\bf R}) ,
\label{dynamic_scaling}
\end{equation}
where the dependence of the line width on the length scales has now
been made explicit in its argument.  Inserting the above scaling laws,
Eq.(\ref{static_scaling}) and Eq.(\ref{dynamic_scaling}) and the
scaling of the vertex functions
\begin{eqnarray}
K^{\beta \gamma }_{\alpha} ({\bf k}, {\bf q}; \xi,q_{_D}, q_{_A})
 = q^4 \hat{K}^{\beta \gamma }_{\alpha} (\bbox{\rho}, {\hat{\bf q}}; {\bf R})
\end{eqnarray}
we get
\begin{eqnarray}
\gamma_{\alpha}(\nu; {\bf R}) & = &
\int_0^{\infty} \! d \rho \, \rho^2 \int_0^{\pi} \!
d \eta \sin\eta \int_0^{2 \pi} d \sigma
\frac{\hat{K}^{\beta \gamma }_{\alpha} (\bbox{\rho},{\hat{\bf q}}; {\bf R})}
{\hat{\chi}_{\alpha}({\hat{\bf q}}; {R})}
\nonumber \\
 &\times&
\frac{ \hat{\chi}_{\beta}(\bbox{\rho}; {\bf R}) \,
       \hat{\chi}_{\gamma}(\bbox{\rho}_-; {\bf R}) }
     {\rho^{5/2} \gamma_{\beta}(\eta, {\bf R}/\rho) +
      \rho_{-}^{5/2} \gamma_{\gamma}(\eta_{-},{\bf R}/\rho_{-})}.
\end{eqnarray}
where $\bbox{\rho} = {\bf k}/q$, $\bbox{\rho}_- = ({\bf k} - {\bf
  q})/q$, and the first argument in the scaling functions for the line
width now gives the azimuthal angle between the wave vector and the
easy axis of magnetization.  The non universal scale factor
\begin{eqnarray}
A =  \sqrt{ \frac{4 J k_B T  V^2}{N^2 (2 \pi)^3} },
\end{eqnarray}
and the dynamic exponent for the isotropic Heisenberg ferromagnet
\begin{eqnarray}
z & = & \frac{5}{2} .
\end{eqnarray}

Here we have introduced polar coordinates such that
\begin{eqnarray}
{\hat{\bf q}} & = &
\left(
\begin{array}{c}
\cos{\mu} \, \sin{\nu}  \\
\sin{\mu} \, \sin{\nu}  \\
\cos{\nu}
\end{array}
\right) , \\
\bbox{\rho} & = &
\left(
\begin{array}{c}
\rho \, \cos{\sigma} \, \sin{\eta}  \\
\rho \, \sin{\sigma} \, \sin{\eta}  \\
\rho \, \cos{\eta}
\end{array}
\right) , \\
\bbox{\rho}_{-} & = &
\left(
\begin{array}{c}
\rho_{-} \, \cos{\sigma_{-}} \, \sin{\eta_{-}}  \\
\rho_{-} \, \sin{\sigma_{-}} \, \sin{\eta_{-}}  \\
\rho_{-} \, \cos{\eta_{-}}
\end{array}
\right) , \\
\end{eqnarray}
with
\begin{eqnarray*}
\rho_{-}    & = &
 \sqrt{ 1 + \rho^2 - 2
\rho (\sin \nu  \sin \eta  \cos (\mu - \sigma ) +
\cos \nu  \cos \eta ) } , \\
\sigma_{-}  & = &  \arctan \left(
\frac{\sin \mu  \sin \nu - \rho  \sin \sigma  \sin \eta }{\cos \mu
\sin\nu - \rho  \cos \sigma  \sin \eta } \right) , \\
\eta_{-}    & = &  \arccos
\left( \frac{\cos \nu - \rho  \cos \eta}{\rho_-} \right).
\end{eqnarray*}

\subsection{Solution of the mode coupling equations}
\label{solution_mc}

Now we are going to discuss the solution of the mode coupling
equations derived in the preceding section. Since the general case
including uniaxial anisotropy as well as dipolar interaction is quite
complicated we will first shortly review the limiting cases of a)
uniaxial and b) dipolar anisotropy, respectively. Since both cases are
discussed in detail in the literature
\cite{frey-schwabl:94,bagnuls-joukoff_piette:75} we will be rather
brief and mention only those aspects which will be relevant for the
subsequent discussion.

\paragraph{Uniaxial limit:} The equations of motion in the uniaxial limit
reduce to
\begin{eqnarray}
\dot{S}^x_{\bf q}
 & = & - 2 \int_{{\bf k}} J  (q^2 - 2 {\bf q} \cdot {\bf k} -m)
 \, S^y_{\bf k}  S^z_{{\bf q}-{\bf k}},  \\
\dot{S}^y_{\bf q}
 & = & - 2 \int_{{\bf k}} J \,  (q^2 - 2 {\bf q} \cdot {\bf k} -m)
 \,  S^x_{\bf k}  S^z_{{\bf q}-{\bf k}},  \\
\dot{S}^z_{\bf q} & = & -  2 \int_{{\bf k}} J \, (q^2 - 2 {\bf q}
 \cdot {\bf k} ) \,  S^x_{\bf k} S^y_{{\bf q}-{\bf k}}.
\end{eqnarray}
The eigenmode analysis shows that the mode coupling equations (in Lorentzian
approximation) become diagonal in the spin fluctuations $s_\alpha$
perpendicular and parallel to the easy axis of magnetization ($\alpha = \perp,
\parallel$)
\begin{eqnarray}
\Gamma_{\perp}({\bf q})& = &
\frac{4 k_B T J^2}{\chi_{\parallel}({\bf q})}
\int_{{\bf k}}
\frac{(q^2 - 2 {\bf q} \cdot {\bf k} - \xi^{-2})^2}
     {\Gamma_{\perp}({\bf k}) + \Gamma_{\parallel}({\bf q}\!-\!{\bf k})},
\label{mc_uniaxial_1} \\
\Gamma_{\parallel}({\bf q})& = &
\frac{4 k_B T J^2}{\chi_{\parallel}({\bf q})}
\int_{{\bf k}}
\frac{(q^2 - 2 {\bf q} \cdot {\bf k})^2}
     {\Gamma_{\perp}({\bf k})+ \Gamma_{\perp}({\bf q}\!-\!{\bf k})}.
\label{mc_uniaxial_2}
\end{eqnarray}
The equations (\ref{mc_uniaxial_1}) and (\ref{mc_uniaxial_2}) obey the
dynamical scaling laws \cite{ferrell-etal:67a,ferrell-etal:67b}
\begin{equation}
\Gamma_{\alpha} (q; \xi, q_{_A}) =
A  \, q^{{z}} \,  \gamma_{\alpha }(x,z) ,
\end{equation}
with the following mode coupling equations for the scaling functions of the
line widths
\begin{eqnarray}
\gamma_{\perp}(x,z) & = &
2 \pi \int_0^{\infty} d \rho \,  \int_0^{\pi} d \eta \, \sin\eta
 \enspace K_{\perp}(\rho,\eta; x,z) \nonumber \\
 & & \times \Bigl[ \rho^{\frac{5}{2}} \, \gamma_{\perp}(\frac{x}{\rho},\frac{z}{\rho})
 + \rho_{-}^{\frac{5}{2}} \,
\gamma_{\parallel}(\frac{x}{\rho_{-}},\frac{z}{\rho_{-}}) \Bigr] ^{-1} ,
\label{mc_scal_uniaxial_1} \\
\gamma_{\parallel}(x,z) & = &
2 \pi \int_0^{\infty} d \rho \, \int_0^{\pi} d \eta \, \sin\eta
 \enspace K_{\parallel}(\rho,\eta; x,z) \nonumber \\
 & & \times \Bigl[\rho^{\frac{5}{2}} \, \gamma_{\perp}(\frac{x}{\rho},\frac{z}{\rho})
 + \rho_{-}^{\frac{5}{2}} \,
\gamma_{\perp}(\frac{x}{\rho_{-}},\frac{z}{\rho_{-}}) \Bigr]^{-1} ,
\label{mc_scal_uniaxial_2}
\end{eqnarray}
and the scaled vertices,
\begin{eqnarray}
 K_{\perp}(\rho,\eta; x,z) & = & \frac{\hat{\chi}_{\perp}(\rho,x,z) \,
\hat{\chi}_{\parallel}(\rho_{-},x)}{\hat{\chi}_{\perp}(1,x,z)} \nonumber \\
&& \times  \rho^2 \, (1 - 2 \rho \cos\eta - z^2)^2  ,
\label{vertex_scal_uniaxial_1} \\
 K_{\parallel}(\rho,\eta; x,z) & = & \frac{\hat{\chi}_{\perp}(\rho,x,z) \,
\hat{\chi}_{\perp}(\rho_{-},x)}{\hat{\chi}_{\parallel}(1,x,z)} \nonumber \\
&& \times  \rho^2 \, (1 - 2 \rho \cos\eta)^2  ,
\label{vertex_scal_uniaxial_2}
\end{eqnarray}
the dynamic exponent $z=5/2$ and the non-universal scale
\begin{eqnarray}
A =  \sqrt{\frac{4 J \, k_B \, T \, V^2}{ (2 \pi)^3 N^2}} .
\end{eqnarray}
The crossover of the critical dynamic exponent is contained in the
scaling functions $\gamma_\alpha (x,z)$. The above mode coupling
equations are essentially the same as those derived in
Ref.~\cite{bagnuls-joukoff_piette:75} and their numerical results are
confirmed by our analysis here (see below).

As summarized in table~\ref{table_asymptotic_uniaxial} the mode
coupling equations can be solved analytically in the uniaxial (U) and
isotropic (I) critical (C) and hydrodynamic (H) limiting regions.
These are defined by UC: $z \gg1$, $x \ll 1$; IC: $z \ll 1$, $x \ll
1$; UH: $z \gg x$, $x \gg 1$; IH: $z \ll x$, $x \gg 1$.

\begin{figure}[htb]
  \centerline{\epsfig{figure=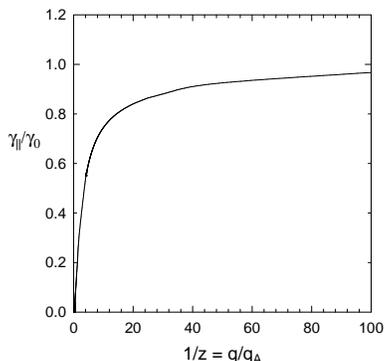,height=5.0cm,clip=}}
\caption{Scaling function for the line width of the spin fluctuations parallel
  to the easy axis of magnetization, $\gamma_{\parallel}(x, z)$, as a
  function of $q/q_{_A} $ at the critical temperature $x=0$ ($T =
  T_c$).}
\label{figure3}
\end{figure}

\begin{figure}[htb]
\centerline{\epsfig{figure=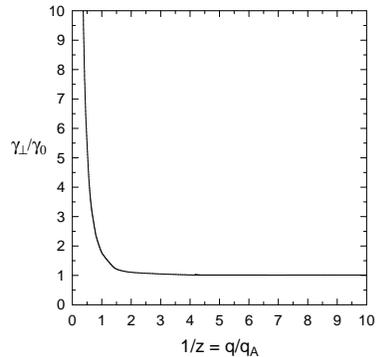,height=5.0cm,clip=}}
\caption{Scaling function for the line width of the spin fluctuations
  perpendicular to the easy axis of magnetization $\gamma_{\perp}(x,
  z)$ as a function of $q/q_{_A} $ at the critical temperature $x=0$
  ($T = T_c$).}
\label{figure4}
\end{figure}

To examine the uniaxial crossover precisely at the critical
temperature, Fig.~\ref{figure3} and Fig.~\ref{figure4} display the
results of the mode coupling theory for the scaling functions for the
spin fluctuations parallel and perpendicular to the easy axis of
magnetization for $T=T_c$ against the wave vector; i.e.\ $z^{-1} =
q/q_{_A}$. We define an effective dynamic exponent by
\begin{equation}
\Gamma_{\alpha}(q;\xi,q_{_A})
\Bigg \vert_{T=T_c} \propto q^{{z}_{{\rm eff},\alpha}},
\qquad \alpha = \parallel , \, \perp .
\end{equation}
The results of the mode coupling theory clearly show that the
crossover from isotropic to uniaxial critical dynamics in the
transverse line width (i.e.\ perpendicular to the easy axis of
magnetization) occurs in the immediate vicinity of the uniaxial
crossover wave vector $q_{_A}$ from the isotropic Heisenberg value
$z_{\rm eff} = 5/2$ to $z_{\rm eff} = 0$.  In contrast, the crossover
in the longitudinal line width (i.e.\ parallel to the easy axis of
magnetization) from an effective exponent $z_{\rm eff} = 5/2$ to
$z_{\rm eff} = 4$ occurs at a wave number {\em larger} than $q_{_A}$
by approximately one order of magnitude. It is interesting to compare
this dynamical shift of the crossover position with the analogous
problem of the crossover from isotropic Heisenberg to isotropic
dipolar dynamics, where the shift of the crossover position in the
critical mode (i.e.\ transverse to the wave vector of the spin
fluctuations) is shifted to wave vectors {\em smaller} than the
corresponding anisotropy scale, the dipolar wave vector $q_{_D}$.

\begin{figure}[htb]
  \centerline{\epsfig{figure=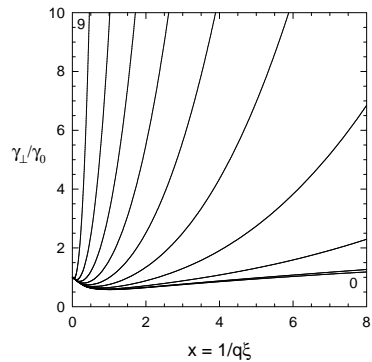,height=5.0cm,clip=}}
\caption{Scaling function for the linewidth of the spin fluctuations
  perpendicular to the easy axis of magnetization $\gamma_{\perp}$ as
  function of $x = 1/ q \xi$ for various angles $\varphi = N \pi/20$,
  with $N = 0, \ldots , 9$.}
\label{figure7}
\end{figure}

\begin{figure}[htb]
  \centerline{\epsfig{figure=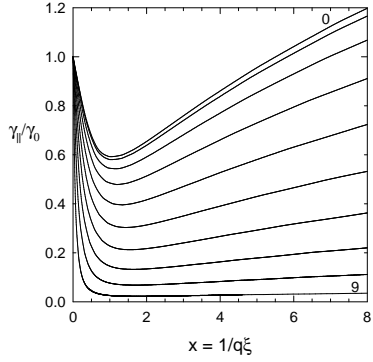,height=5.0cm,clip=}}
\caption{Scaling function for the linewidth of the spin fluctuations
  parallel to the easy axis of magnetization $\gamma_{\parallel}$ as
  function of $x = 1/ q \xi$ for various angles $\varphi = N \pi/20$,
  mit $N = 0, \ldots , 9$.}
\label{figure8}
\end{figure}

The physical content of the two parameter scaling surfaces is
illustrated best by considering cuts for fixed $q_{_A}$ and various
temperatures, since for a given material, $q_{_A}$ is fixed and the
parametrisation by $\varphi$ corresponds to a parametrisation in terms
of the reduced temperature $(T-T_c)/T_c$.  In
Figs. \ref{figure7}-\ref{figure8} the scaling functions versus $x =
1/q\xi$ are displayed for different values of $\varphi = \arctan
(q_{_{A}} \xi) = N \pi / 20$ with $N = 0,1,...,9$. For $\varphi = 0$,
corresponding to vanishing uniaxial anisotropy $q_{_A}$, the scaling
functions coincide with the Resibois-Piette scaling function
\cite{resibois-piette:70}.  If the strength of the uniaxial anisotropy
$q_{_A}$ is finite, the curves for the scaling functions approach the
Resibois-Piette scaling function for small values of the scaling
variable $x$ and deviate therefrom with increasing $x$.  At fixed
scaling variable $x$ and with increasing temperature (increasing
$\varphi$) the value of the scaling function for the spin
fluctuations along the easy axis of magnetization is lowered with
respect to the Resibois-Piette function, whereas it is increased for
the scaling function of the spin fluctuations in the basal plane.

\paragraph{Dipolar limit:}
As summarized in Table~\ref{table_asymptotic_dipolar} the mode
coupling equations in the pure dipolar limit can be solved
analytically in the dipolar (D) and isotropic (I) critical (C) and
hydrodynamic (H) limiting regions. These are defined by DC: $y \gg1$,
$x \ll 1$; IC: $y \ll 1$, $x \ll 1$; DH: $y \gg x$, $x \gg 1$; IH: $y
\ll x$, $x \gg 1$.

Concerning the critical dynamical exponent one finds for the
longitudinal line width a crossover from $z = {5 / 2}$ in the
isotropic critical region to $z = 0$ in the dipolar critical region,
whereas for the transverse line width the crossover is from $z = {5 /
  2}$ to $z = 2$. The precise position of this crossover can only be
determined numerically.

A plot of the transverse and longitudinal
scaling functions $\gamma^T(x,y)$ and $\gamma^L (x,y)$ can be found in
Ref.~\cite{frey-schwabl:94}. For the dipolar crossover precisely at
the Curie point the crossover from the isotropic Heisenberg to dipolar
critical dynamics in the transverse line width occurs at a wave
number, which is almost one order of magnitude smaller than the static
crossover wave vector $q_{_D}$.  The crossover of the longitudinal
width, from $z=2.5$ to $z=0$, is more pronounced and occurs in the
intermediate vicinity of $q_{_D}$.  The reason for the different
location of the dynamic crossover is mainly due to the fact that it is
primarily the longitudinal static susceptibility which shows a
crossover due to the dipolar interaction.  Since the change in the
static critical exponents is numerically small the transverse static
susceptibility is nearly the same as for ferromagnets without dipolar
interaction.  Hence the crossover in the transverse width is purely a
dynamical crossover, whereas the crossover of the longitudinal width
being proportional to the inverse static longitudinal susceptibility
is enhanced by the static crossover.

\begin{figure}[htb]
\centerline{\epsfig{figure=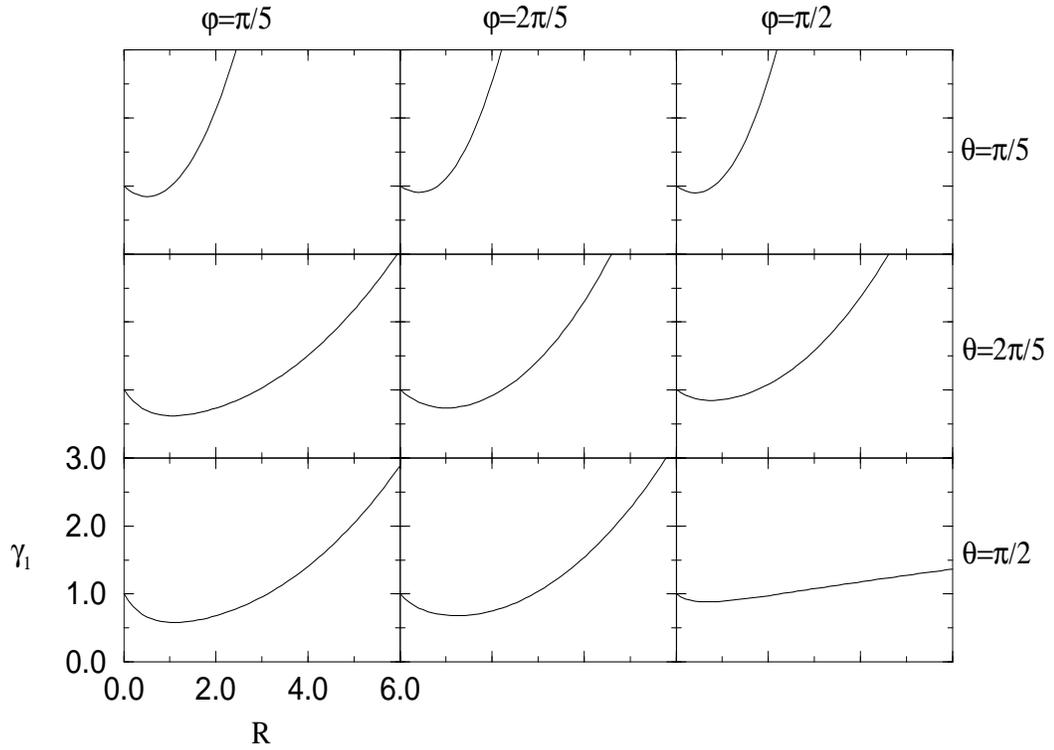,height=10cm,width=14cm}}
\vspace{15pt}
\caption{Scaling function $\gamma_1$ as a function of $R$ for various angles
  $\theta$ and $\varphi$. Each graph for fixed $\theta$ and $\varphi$ contains a set of
  curves parametrized by the angle $\nu$ of the wave vector with
  respect to the $z$-axis: $\nu = N \pi/8$ with $N = 0, \cdots, 4$.}
\label{figure12}
\end{figure}

\begin{figure}[htb]
  \centerline{\epsfig{figure=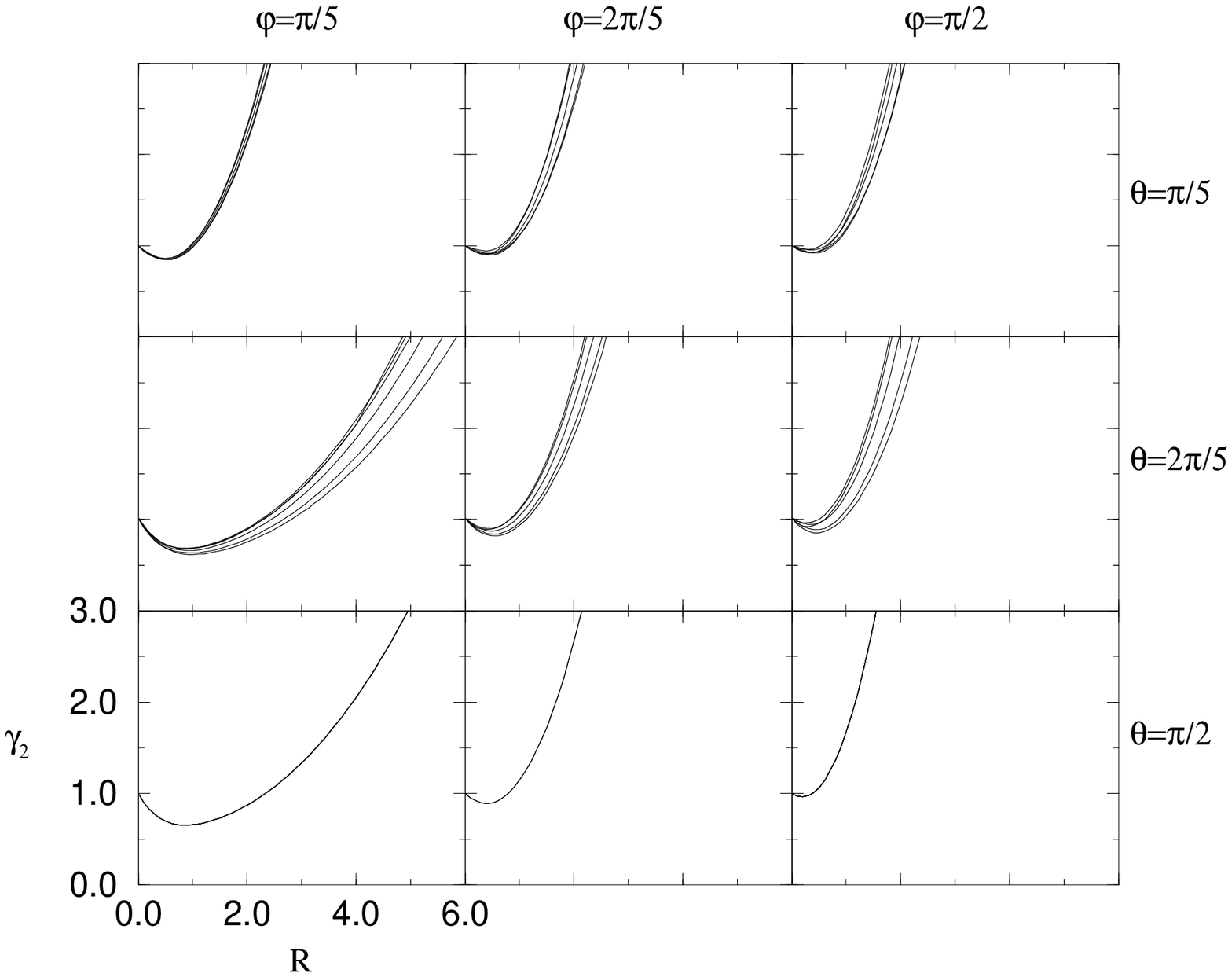,height=10cm,width=14cm}}
  \vspace{15pt}
\caption{Scaling function $\gamma_2$ as a function of $R$ for various angles
  $\theta$ and $\varphi$. Each graph for fixed $\theta$ and 
  $\varphi$ contains a set of
  curves parametrized by the angle $\nu$ of the wave vector with
  respect to the $z$-axis: $\nu = N \pi/8$ with $N = 0, \cdots, 4$.}
\label{figure13}
\end{figure}

\begin{figure}[htb]
\centerline{\epsfig{figure=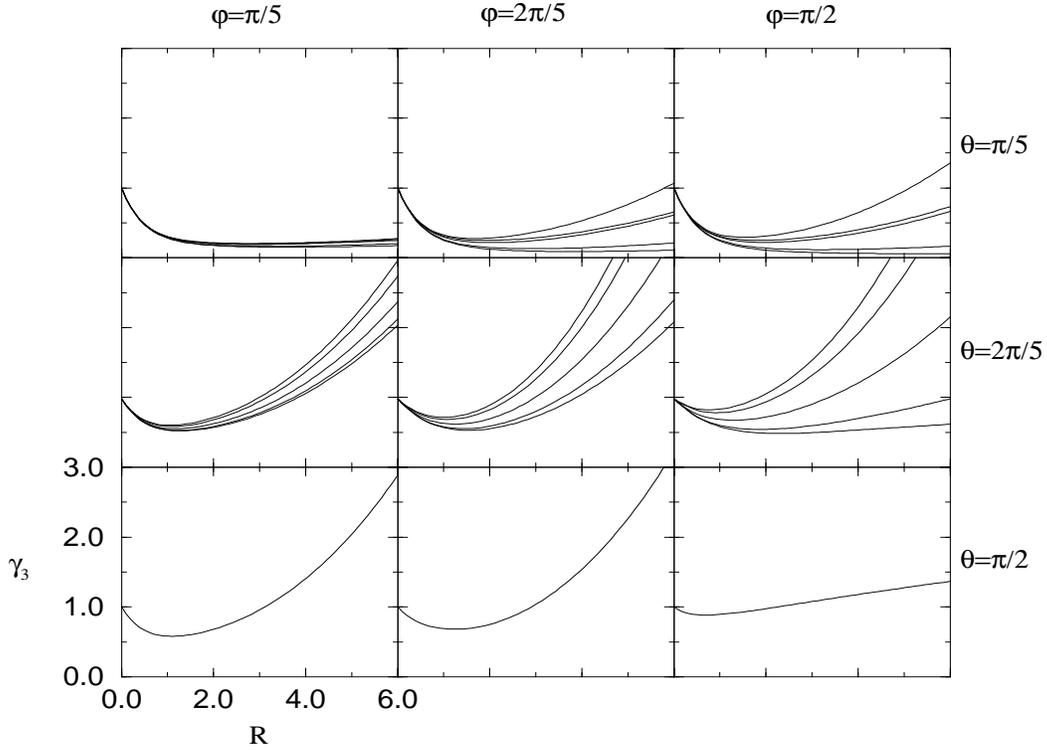,height=10cm,width=14cm}}
\vspace{15pt}
\caption{Scaling function $\gamma_3$ as a function of $R$ for various angles
  $\theta$ and $\varphi$.
  Each graph for fixed $\theta$ and $\varphi$ contains a set of
  curves parametrized by the angle $\nu$ of the wave vector with
  respect to the $z$-axis: $\nu = N \pi/8$ with $N = 0, \cdots, 4$.}
\label{figure14}
\end{figure}

\paragraph{General case:} For nonvanishing dipolar interaction and uniaxial
anisotropy the mode coupling equations for the scaling functions of
the spin fluctuations $s_\alpha ({\bf q})$ are given by
\begin{eqnarray}
\gamma_{\alpha}(\nu; {\bf R}) & = &
\int_0^{\infty} d \rho \, \rho^2 \int_0^{\pi} d \eta \sin\eta \int_0^{2 \pi}
d \sigma  \nonumber \\
&\times&
{\hat K}^{\beta \gamma }_{\alpha} ({\rho},{\hat{\bf q}}; {\bf R}) \,
\frac{\hat{\chi}_{\beta}(\bbox{\rho}; {\bf R}) \,
      \hat{\chi}_{\gamma}(\bbox{\rho}_{-}; {\bf R})}
     {\hat{\chi}_{\alpha}({\hat{\bf q}}; {\bf R})}  \nonumber \\
&\times&
   \frac{1}
        {\rho^{\frac{5}{2}}
         \gamma_{\beta}(\eta; \frac{R}{\rho}, \varphi, \theta)
       + \rho_{-}^{\frac{5}{2}}
         \gamma_{\gamma}(\eta_{-}; \frac{R}{\rho_{-}}, \varphi, \theta)},
\label{mc_scal_general}
\end{eqnarray}
where we have introduced polar coordinates
\begin{eqnarray}
{\bf R} & = &
\left( 
\begin{array}{c}
x \\
y \\
z \\
\end{array} \right)
\, = \, \left( \begin{array}{c}
R \, \cos \varphi \, \sin \theta \\
R \, \sin \varphi \, \sin \theta \\
R \, \cos \theta \\
\end{array} \right) ,
\end{eqnarray}
with
\begin{eqnarray}
R & \in & [0; \infty [ \qquad \varphi \, \in \, [0; \frac{\pi}{2} ] \qquad
\theta \, \in \,  [0; \frac{\pi}{2} ] . \nonumber
\end{eqnarray}
The mode coupling equations can be solved analytically only in some
limiting cases.  The results for these limiting cases are given table
\ref{table_asymptotic_general}. In all other cases one has to solve
the mode coupling equations numerically. For this it is best to write
the scaling variables in spherical coordinates. The results for the
three scaling functions are plotted in
Figs.~\ref{figure12}--\ref{figure14}. In these figures we have plotted
a table for each of the scaling functions, where in each element of
the table the scaling function is shown as a function of the radial
scaling variable $R$ for a set of angles of the wave vector with
respect to the $z$-axis. The three colums (from left to right)
correspond to the angles $\varphi = \pi/5$, $2\pi/5$ and $\pi/2$ and
the rows (from top to bottom) correspond to $\theta = \pi/5$, $2\pi/5$
and $\pi/2$.  Note that $\tan \varphi = q_{_D} \xi$ and $\tan \theta =
\sqrt{\xi^{-2}+q_{_D}^2}/q_{_A}$.

The continuous crossover from isotropic to uniaxial or dipolar
behavior can then be easily read off from
Figs.~\ref{figure12}--\ref{figure14}.
\begin{itemize}
\item For $\varphi \rightarrow 0$ and $\theta \rightarrow \pi/2$ all
  three scaling functions become identical to the Resibois-Piette
  scaling function for the isotropic Heisenberg ferromagnet.
\item In the uniaxial limit $\varphi \rightarrow 0$ (corresponding to
  $q_{_D} \rightarrow 0$) the first two scaling functions become identical.
\item Variation of the anisotropy strength $q_{_A}$ in the weak dipolar
  limit ($q_{_D} \xi \ll 1$) (i.e., increasing $\theta$ from $0$ to
  $\pi/2$ for small values of $\varphi$) gives the transition from the
  Resibois-Piette function to the uniaxial limit discussed above.
  $\gamma_1$ and $\gamma_2$ become identical to the scaling function
  for the uncritical transverse modes whereas $\gamma_3$ becomes the
  scaling function for the fluctuations along the easy axis.
\item For small uniaxial anisotropy ($q_{_A} \ll \sqrt{\xi^{-2} + q_{_D}^2}$
  or equivalently $\theta$ close to $\pi/2$) one recovers the dipolar
  limit where $\gamma_1$ and $\gamma_3$ reduce to the two equivalent
  transverse modes, and $\gamma_2$ becomes the uncritical longitudinal
  mode.
\item Upon going along the diagonal from top-left to bottom-right in
  Figs.~\ref{figure12}--\ref{figure14} one crosses over from the
  uniaxial to the dipolar limit and $\gamma_1$ and $\gamma_3$ become
  more and more similar. The scaling function $\gamma_2$ makes a
  continuous crossover from the scaling function for the mode parallel
  to the easy axis (critical mode) in the uniaxial case to the scaling
  function of the longitudinal mode (uncritical mode) in the dipolar
  limit.
\item The first mode $\gamma_1$ does not show any dependence on the
  orientation $\nu$ of the wave vector with respect to the $z$-axis.
  The reason is that this mode is by construction always perpendicular
  to both the $z$-axis and the wave vector ${\bf q}$ and hence
  idependent from the angle between these two axes.
\item The second and third mode show no $\nu$-dependence neither in
  the dipolar nor in the uniaxial limit. The dependence on $\nu$ is
  strongest close to $\theta \approx \pi/4$ and $\varphi \approx
  \pi/4$. It is most pronounced for the third mode. This is due to the
  strong $\nu$-dependence of this mode in the hydrodynamic limit (see
  table \ref{table_asymptotic_general}). Whereas the scaling behavior
  of $\gamma_1$ and $\gamma_2$ does not depend on the orientation of
  the wave vector with respect to the $z$-axis in this limit.
\item The effective dynamic exponents $z_{\rm eff}$ become zero for all three modes 
  in the case $q_{_{D}}\neq 0$, except for $\gamma_{3}$ in the case $\nu=\pi/2$. 
\end{itemize}

\end{multicols}
\widetext

\begin{multicols}{2}

\narrowtext

\paragraph{Gd -- no magnetocrystalline anisotropy:} 
As we have already mentioned in the introduction Gd is a very
interesting rare earth matrerial because it is an S-state ion with a
large localized magnetic moment. Because of this it should have a very
small magnetocrystalline anisotropy and be a much better model system
for an isotropic Heisenberg ferromagnet than materials like Fe, EuO or
EuS.  Contrary to this expectation, however, experimental observations
teach us that Gd has an easy axis which coincides with the hexagonal
axis of its hcp structure. In the following section we will identify
the dipole--dipole interaction in conjunction with the lattice
structure as the source of this easy axis anisotropy. Then we can
explicitely calculate the ratio of the two crossover wave vectors,
\begin{eqnarray}
q_{_D} / q_{_A} = 7.8738 \, ,
\end{eqnarray}
and thus reduce the number of material parameters by one. This allows
us to present our theoretical results for the line widths of the
correlation functions in the same way as for the two limiting cases of
the cubic dipolar and purely uniaxial systems.

\begin{figure}[htb]
\centerline{\epsfig{figure=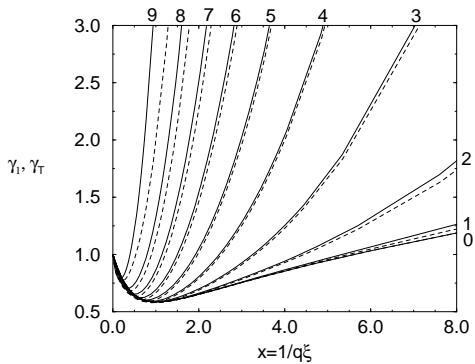,height=5.0cm}}
\vspace{15pt}
\caption{Comparison of the scaling function $\gamma_1$ at $\nu = 0$ 
  of the hcp dipolar model (solid lines) with the transverse scaling
  function $\gamma_T$ (dashed lines) of the cubic dipolar model.  The
  scaling functions are given as a function of $x=1/q \xi$ for various
  values of $\varphi = N \pi /20$ with $N = 0, ..., 9$.}
\label{fig:comp1}
\end{figure}

\begin{figure}[htb]
\centerline{\epsfig{figure=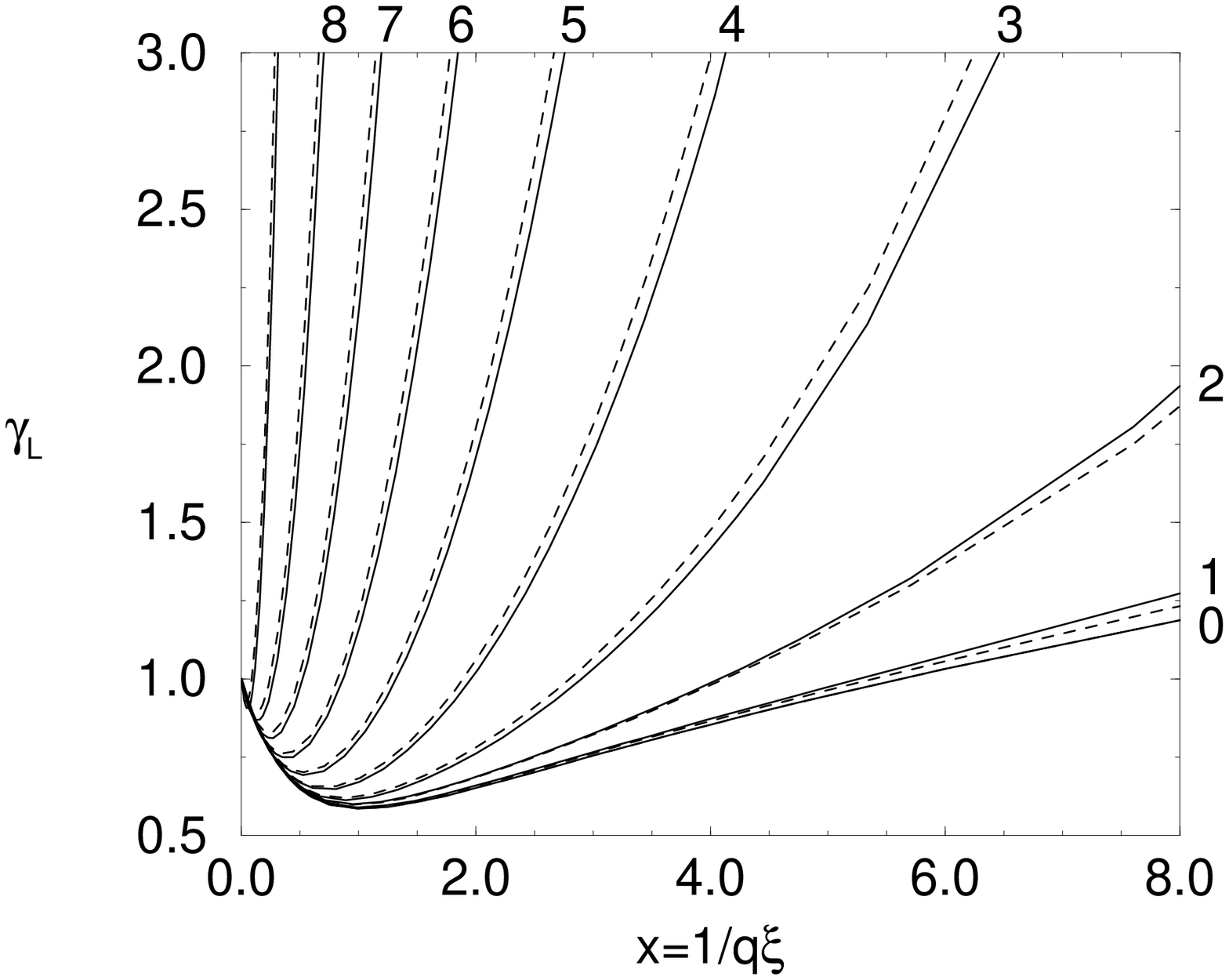,height=5.0cm}}
\vspace{15pt}
\caption{Comparison of the scaling function $\gamma_2$ at $\nu = 0$ 
  of the hcp dipolar model (solid lines) with the longitudinal scaling
  function $\gamma_L$ (dashed lines) of the cubic dipolar model.  The
  scaling functions are given as a function of $x=1/q \xi$ for various
  values of $\varphi = N \pi /20$ with $N = 0, ..., 9$.}
\label{fig:comp2}
\end{figure}

\begin{figure}[htb]
\centerline{\epsfig{figure=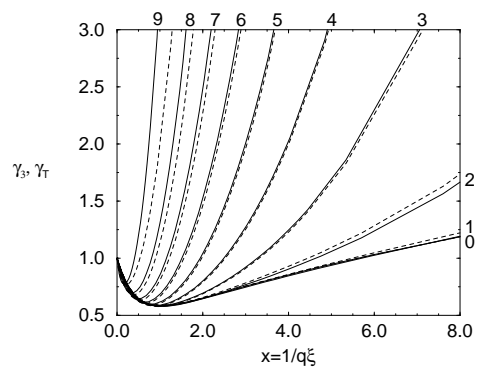,height=5.0cm}}
\vspace{15pt}
\caption{Comparison of the scaling function $\gamma_3$ at $\nu = 0$ 
  of the hcp dipolar model (solid lines) with the transverse scaling
  function $\gamma_T$ (dashed lines) of the cubic dipolar model.  The
  scaling functions are given as a function of $x=1/q \xi$ for various
  values of $\varphi = N \pi /20$ with $N = 0, ..., 9$.}
\label{fig:comp3}
\end{figure}

In Figures \ref{fig:comp1}--\ref{fig:comp3} we compare the scaling
functions for the line widths of the three modes $\gamma_1$,
$\gamma_2$ and $\gamma_3$ for the hcp dipolar model with the
corresponding results for the cubic dipolar model.
Since the dipolar interaction is the dominant anisotropy the
deviations of the results for the hcp from the cubic model are small.
For $\gamma_1$ and $\gamma_3$ we find that the relaxation times of the
hcp system are reduced as compared to the cubic model. This seems
plausible since -- as discussed in section III -- the induced uniaxial
anisotropy supresses some of the critical fluctuations. This has to be
contrasted with the behavior of $\gamma_2$, where the corresponding
longitudinal mode in the cubic dipolar system is alread an uncritical
mode. We find that the result for $\gamma_2$ is almost identical to
the longitudinal scaling function of the cubic dipolar system.

\begin{figure}[htb]
\centerline{\epsfig{figure=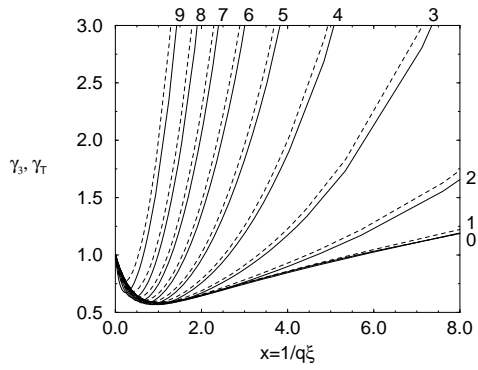,height=5.0cm}}
\vspace{15pt}
\caption{Comparison of the scaling function $\gamma_3$ at $\nu = \pi / 2$ 
  of the hcp dipolar model (solid lines) with the transverse scaling
  function $\gamma_T$ (dashed lines) of the cubic dipolar model.  The
  scaling functions are given as a function of $x=1/q \xi$ for various
  values of $\varphi = N \pi /20$ with $N = 0, ..., 9$.}
\label{fig:comp3_n4}
\end{figure}

The case $\nu = \pi/2$ is special since here the third mode in the hcp
model becomes critical (see section III). As a consequence there is no
supression of critical fluctuations and one does not expect that
dynamics is slowed down with respect to the cubic case. Indeed, as can
be inferred from Fig.~\ref{fig:comp3_n4}, the mode becomes even faster
than in the cubic case.

\begin{figure}[htb]
\centerline{\epsfig{figure=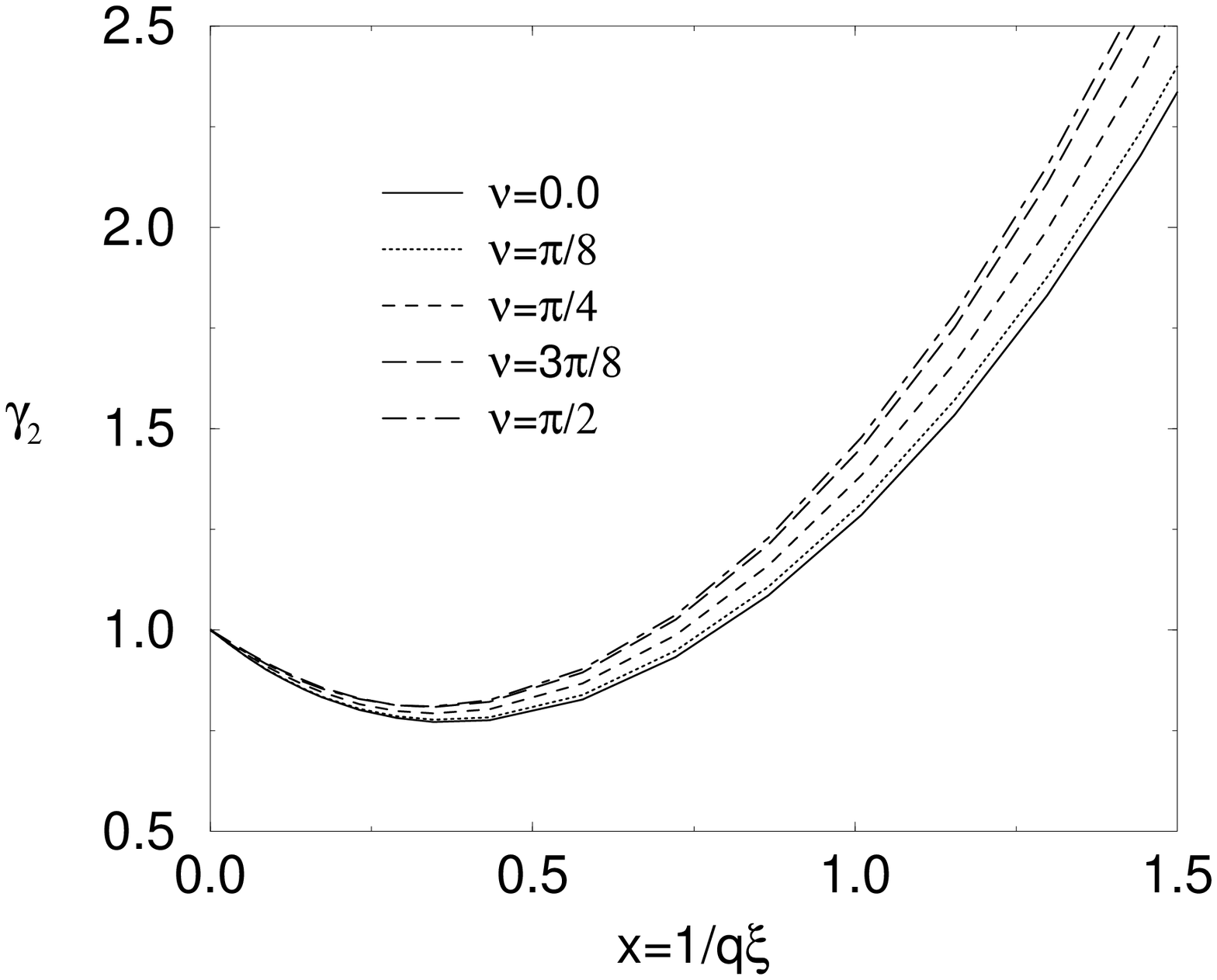,height=5.0cm}}
\vspace{15pt}
\caption{Scaling function $\gamma_2$ for the hcp
  dipolar model as a function of $x=1/q\xi$ for fixed values $q_{_D} \xi
  = 1.0$ and $q_{_A} \xi = 1.0$ and a series of angles $\nu = N \pi / 8$
  with $N = 0,1, \cdots, 4$ indicated in the graph.}
\label{fig:nueseq2}
\end{figure}

\begin{figure}[htb]
\centerline{\epsfig{figure=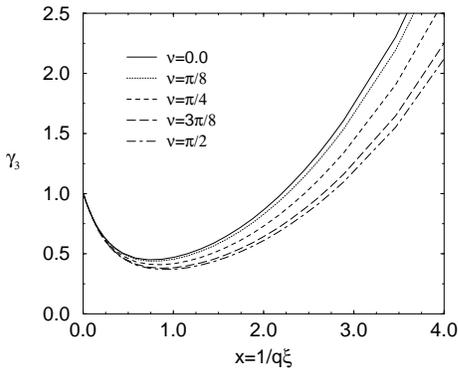,height=5.0cm}}
\vspace{15pt}
\caption{Scaling function $\gamma_3$
  for the hcp dipolar model as a function of $x=1/q\xi$ for fixed
  values $q_{_D} \xi = 1.0$ and $q_{_A} \xi = 1.0$ and a series of angles
  $\nu = N \pi / 8$ with $N = 0,1, \cdots, 4$ indicated in the graph.}
\label{fig:nueseq3}
\end{figure}

The dependence on the relative orientation of the
$c$-axis and the wave vector is not very pronounced. There is actually
no $\nu$-dependence for $\gamma_1$ since this mode is by construction
always perpendicular to the $c$-axis. In
Figs.~\ref{fig:nueseq2}-\ref{fig:nueseq3} we try to visualize the
$\nu$-dependence of $\gamma_2$ and $\gamma_3$ at a set of parameters
$q_{_D} \xi = 1.0$ and $q_{_A} \xi = 1.0$.  These parameters do not
correspond to Gd, because as we have already seen in
Figs.~\ref{figure12}-\ref{figure14}, the angle dependence is largest
when $q_{_A}$ and $q_{_D}$ are comparable; hence it is quite small for
the hcp dipolar system Gd.


\begin{figure}[htb]
\centerline{\epsfig{figure=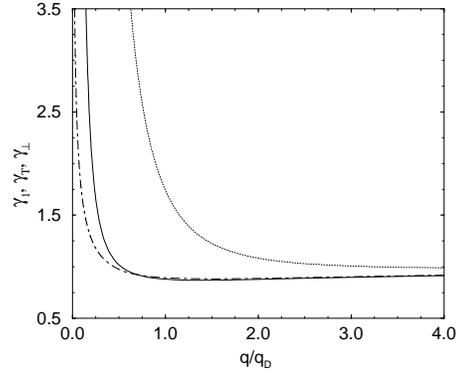,height=5.0cm}}
\vspace{15pt}
\caption{Scaling function $\gamma_1$ at the critical temperature
  as a function of $q/q_{_D}$ (solid line) compared with the
  corresponding transverse scaling function $\gamma_{_{T}}$ of the cubic 
  dipolar system (dot-dashed line) and the hard axis scaling 
  function $\gamma_{_{\perp}}$of the
  uniaxial system (dotted line).  Note that in the uniaxial case the
  scaling function is plotted as a function of $q/q_{_A}$ but as a
  function of $q/q_{_D}$ for the dipolar cases.}
\label{fig:Mode1Tc}
\end{figure}

\begin{figure}[htb]
\centerline{\epsfig{figure=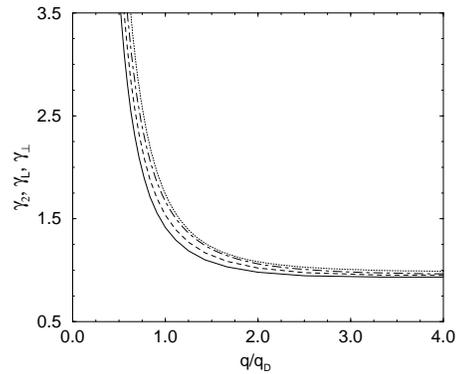,height=5.0cm}}
\vspace{15pt}
\caption{Scaling function $\gamma_2$ at the critical temperature
  as a function of $q/q_{_D}$ for $\nu = 0$ (solid line) and $\nu =
  \pi/2$ (dashed line) compared with the corresponding longitudinal 
  scaling function $\gamma_{_{L}}$ of the cubic dipolar system 
  (dot-dashed line) and the hard axis scaling function $\gamma_{_{\perp}}$
  of the uniaxial system (dotted line).  Note that in the
  uniaxial case the scaling function is plotted as a function of
  $q/q_{_A}$ but as a function of $q/q_{_D}$ for the dipolar cases.}
\label{fig:Mode2Tc}
\end{figure}

\begin{figure}[htb]
\centerline{\epsfig{figure=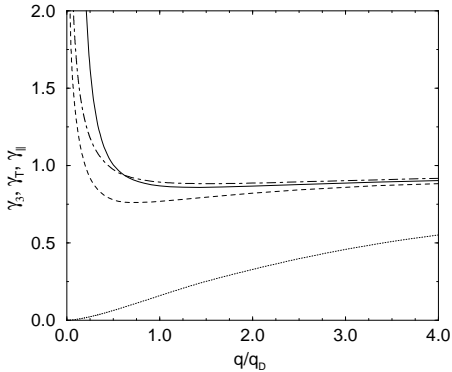,height=5.0cm}}
\vspace{15pt}
\caption{Scaling function $\gamma_3$ at the critical temperature
  as a function of $q/q_{_D}$ for $\nu = 0$ (solid line) and $\nu =
  \pi/2$ (dashed line) compared with the corresponding transverse scaling 
  function $\gamma_{_{T}}$ of the cubic dipolar system (dot-dashed line) 
  and the soft axis scaling function $\gamma_{_{\parallel}}$ of the 
  uniaxial system (dotted line).  Note that in the
  uniaxial case the scaling function is plotted as a function of
  $q/q_{_A}$ but as a function of $q/q_{_D}$ for the dipolar cases.}
\label{fig:Mode3Tc}
\end{figure}

Finally, let us discuss the behavior of the line width right at the
critical temperature. Figs. \ref{fig:Mode1Tc}--\ref{fig:Mode3Tc} show the 
scaling functions of the three modes in the hcp system compared 
with the corresponding results for the cubic dipolar and the uniaxial 
system.  

Again, we find that for the parameter values of Gd there is little
difference between the results for the cubic and the hcp dipolar
system (but a huge difference to the results of a uniaxial model with
no dipolar interaction). The most notable effect is that the crossover
wave vector which marks the deviation from isotropic Heisenberg
behavior (with a dynamic exponent $z=5/2$) is shifted to larger values
for the first mode and to smaller values for the second and third
mode. Then there is also a clearly visible dependence on the
orientation of the wave vector with respect to the $c$--axis for the
third mode. 

\section{Comparison with experimental data}
\label{experiment}

The critical dynamics of Gd has been investigated experimentally
almost exclusively by several nuclear, i.e., hyperfine interaction
(HFI), methods.  The application of nuclear techniques to study
critical phenomena in magnets has recently been reviewed by Hohenemser
et\ al.  \cite{hohenemser-rosov-kleinhammes:89}.  All of the hyperfine
interaction methods are local probes which are related to a wave
vector integral of the spin correlation functions. As such they offer
a complement to neutron scattering. Dynamic studies, using hyperfine
interaction probes, utilize the process of nuclear relaxation produced
by time--dependent hyperfine interaction fields, which reflect
fluctuations of the surrounding electronic magnetic moments.

The hyperfine interaction stems from the magnetic interaction of the electrons
with the magnetic field produced by the nucleus. The hyperfine interaction
of a nucleus with spin ${\bf I}$, $g$--factor $g_N$ and mass $m_N$ with one
of the surrounding electrons with spin ${\bf S}$ and orbital momentum
${\bf L}$ can be written in the form~\cite{frey-schwabl:94}
\begin{eqnarray}
  H_{\rm hyp} &=&
  \frac{Z e_0^2 g_N}{2 m_N m c^2}
  \Biggl[
  \frac{1}{r^3} {\bf I} \cdot {\bf L} +
  \frac{8 \pi}{3} \delta^{(3)} ({\bf x}) \, {\bf I} \cdot {\bf S} \nonumber \\
  &\quad& - \frac{1}{r^3} {\bf I} \cdot {\bf S}  +
  \frac{3 ({\bf I}\cdot{\bf x}) ({\bf S}\cdot{\bf x})}{r^5}
  \Biggr] \, .
\label{hyperfine}
\end{eqnarray}
The first term represents the interaction of the orbital momentum of
the electron with the nuclear magnetic moment of the nucleus. The
second term is the Fermi contact interaction and the last two terms
represent the dipolar interaction. The Fermi contact interaction is
finite only for electrons having a finite probability density at the
nucleus, i.e. bound $s$-electrons or itinerant electrons. The
Hamiltonian, Eq.~(\ref{hyperfine}), can also be used for the analysis
of spin resonance experiments with muons ($\mu$SR).  However, these do
not have bound electrons and hence the Fermi contact term involves
only conduction electrons and is of the same order of magnitude as the
(residual) dipolar interaction~\cite{denison-graf:79}.

\subsection{$\mu$SR measurement}
As shown in appendix~\ref{appendix} the muon damping rate can be
written as
\begin{equation}
\lambda_{\hat z} = \frac{\pi {\cal D}}{V^2}
\int_{{\bf q}}
\sum_{{\hat \beta} {\hat \gamma}}
\left[
G^{{\hat x} {\hat \beta}}_{\bf q} G^{{\hat x} {\hat \gamma}}_{-{\bf q}} +
G^{{\hat y} {\hat \beta}}_{\bf q} G^{{\hat y} {\hat \gamma}}_{-{\bf q}}
\right]
\Phi^{{\hat \beta} {\hat \gamma}}({\bf q}) ,
\label{muon_relax_rate}
\end{equation}
where we have defined ${\cal D} = \gamma_{\mu}^2 (\mu_0/4 \pi)^2 (g_L
\mu_B)^2$.  The coupling of the muon spin and the spins of the magnet
is described in terms of the coupling matrix $G^{{\hat x} {\hat
    \beta}}_{\bf q}$, which reflects the particular symmetry of the
lattice sites occupied by the muons. Since the most dominant
contribution to the damping rate comes from wave vectors close to the
Brillouin zone center, the peculiar properties of the coupling tensor
$G^{{\hat x} {\hat \beta}}_{\bf q}$ at small values of ${\bf q}$ will
be important.  The coupling tensor is determined by both the type of
interaction between the muon spin and the spins of the magnet and the
location of the muon in the lattice.  As noted above the coupling
contains dipolar interaction as well as a contribution from the Fermi
contact field. Using a decomposition into four orthorhombic
sublattices ($l = 0,1,2,3$) the dipolar interaction can be written in
the form \cite{fujiki:87},
\begin{eqnarray}
D^{\alpha \beta}({\bf q})& = & \sum_{l=0}^3 \sum_i D^{\alpha \beta}
({\bf r}_{i,l}) \, e^{i {\bf q} \cdot {\bf r}_{i,l}} ,
\end{eqnarray}
where ${\bf r}_{i,l} = {\bf i} + {\bf r}_l - {\bf r}_0$ denotes the
position on site $i$ of sublattice $l$ with respect to the position of
the muon ${\bf r}_0$. The lowest-order approximation in ${\bf q}$ of
$D^{\alpha \beta} ({\bf q})$ is given by (see appendix \ref{appendix})
\begin{equation}
D^{\alpha \beta}( {\bf q} \to 0)
= - 4 \pi \left[ \frac{q_{\alpha} q_{\beta}}{q^2} - d_{\alpha}  \right]  .
\end{equation}
where we find for octahedral sites ~\cite{reotier-yaouanc:94}:
\begin{displaymath}
d_x = d_y = 0.3485 \quad \mbox{and} \quad d_z =0.3030,
\end{displaymath}
and for tetrahedral sites
\begin{displaymath}
d_x = d_y = 0.3118 \quad \mbox{and} \quad d_z =0.3764 .
\end{displaymath}
Adding the isotropic contribution from the Fermi contact field one
finds in the limit ${\bf q} \to 0$
\begin{eqnarray}
G^{\alpha \beta}_{{\bf q} \to 0}
& = & - 4 \pi \left[ \frac{q_{\alpha} q_{\beta}}{q^2} -
d_{\alpha}  \right] + n_{\mu}  H_{\mu}  \delta_{\alpha \beta} \nonumber \\
& = & - 4 \pi \left[ \frac{q_{\alpha} q_{\beta}}{q^2} - p_{\alpha}  \right] ,
\end{eqnarray}
with
\begin{equation}
p_{\alpha} = d_{\alpha} + \frac{n_{\mu} H_{\mu} }{4 \pi}.
\end{equation}
With the Fermi contact field $B_{\rm FC} = - 6.98 $~kG at $T = 0$~K
\cite{denison-graf:79}, one gets {$n_{\mu} H_{\mu}/4 \pi = -0.278$}
\cite{reotier-yaouanc:94}, and consequently for octahedral sites
\begin{displaymath}
p_x = p_y = 0.0705 \quad \mbox{and} \quad p_z =0.0250 ,
\end{displaymath}
and for tetrahedral sites
\begin{displaymath}
p_x = p_y = 0.0338 \quad \mbox{and} \quad p_z =0.0984 .
\end{displaymath}
Before comparing the theoretical result with experiments one has to
find a transformation between the reference frames used in the
theoretical analysis and the experimental setup, respectively. In
Eq.~\ref{muon_relax_rate} (i.e.  in the experimental setup) one uses a
reference frame $(\hat{x},\hat{y},\hat{z})$ which is fixed with
respect to the initial polarization of the muon beam, which is choosen
to be along the $\hat{z}$-axis.  The reference frame used in the
theoretical analysis, $(x,y,z)$ was chosen such that the $z$-axis
coincides with the easy axis of magnetization.  (see
Fig.~\ref{reference_frame}).  The transformation rules are given by
\begin{eqnarray}
\hat{x} & = & x , \\
\hat{y} & = & y \, \cos \alpha  + z \, \sin \alpha , \\
\hat{z} & = & - y \, \sin \alpha + z \, \cos \alpha ,
\end{eqnarray}
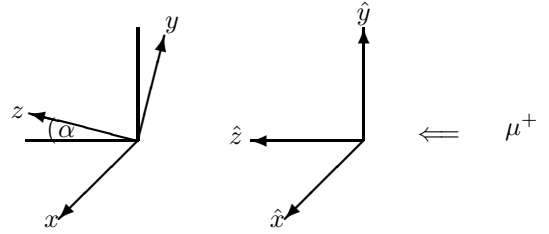
\begin{figure}[htb]
\unitlength0.5cm
\begin{picture}(16,7)
\thicklines
\put(11,3){\vector(-1,0){3}}
\put(11,3){\vector(-1,-1){2.1}}
\put(11,3){\vector(0,1){3}}
\put(5,3){\vector(-4,1){2.9}}
\put(5,3){\vector(-1,-1){2.1}}
\put(5,3){\vector(1,4){0.7}}
\put(5,3){\line(-1,0){3}}
\put(5,3){\line(0,1){3}}
\put(8.7,0.7){\makebox(0,0)[b]{$\hat{x}$}}
\put(11,6.2){\makebox(0,0)[b]{$\hat{y}$}}
\put(7.6,2.9){\makebox(0,0)[b]{$\hat{z}$}}
\put(2.7,0.7){\makebox(0,0)[b]{$x$}}
\put(5.9,6.0){\makebox(0,0)[b]{$y$}}
\put(1.8,3.6){\makebox(0,0)[b]{$z$}}
\put(2.7,3.1){\makebox(0,0)[b]{(}}
\put(3.1,3.1){\makebox(0,0)[b]{$\alpha$}}
\put(13,3){\makebox(0,0)[b]{$\Longleftarrow$}}
\put(15.2,3.15){\makebox(0,0)[b]{$\mu^{+}$}}
\end{picture}
\caption{Reference frame $(\hat{x},\hat{y},\hat{z})$ for $\mu$SR
  measurements, which is rotated by $\alpha$ around the $x$--axis with
  respect to the reference frame $(x,y,z)$. Here $z$ is the direction of
  the easy axis of magnetization and ${\hat z}$ is the direction of
  the initial polarization of the muon beam.}
\label{reference_frame}
\end{figure}

\end{multicols}
\widetext

Then the coupling tensor in the experimental reference frame reads
\begin{eqnarray}
G^{\hat{x} \hat{x}}({\bf q} \to 0 )
& = & - 4 \pi \left[ \frac{q_x^2}{q^2} - p_x \right] , \\
G^{\hat{x} \hat{y}}({\bf q} \to 0 ) & = &
 - 4 \pi \left[ \frac{q_x  q_y}{q^2}
\cos \alpha + \frac{q_x  q_z}{q^2} \sin \alpha - p_x \right] , \\
G^{\hat{x} \hat{z}}({\bf q} \to 0 ) & = &
- 4 \pi \left[- \frac{q_x \ q_y}{q^2}
\sin \alpha + \frac{q_x \ q_z}{q^2} \cos \alpha - p_x \right] , \\
G^{\hat{y} \hat{x}}({\bf q} \to 0 ) & = &
 - 4 \pi \left[ \frac{q_x  q_y}{q^2}
\cos \alpha + \frac{q_x  q_z}{q^2} \sin \alpha - p_y  \cos \alpha
- p_z  \sin \alpha \right] , \\
G^{\hat{y} \hat{y}}({\bf q} \to 0 ) & = & - 4 \pi \Biggl[ \frac{q_y^2}{q^2}
 \cos^2 \alpha + 2 \frac{q_y  q_z}{q^2}  \sin \alpha  \cos \alpha +
\frac{q_z^2}{q^2}  \sin^2 \alpha - p_y  \cos \alpha - p_z  \sin \alpha
\Biggr] , \\
G^{\hat{y} \hat{z}}({\bf q} \to 0 ) & = & - 4 \pi \Biggl[
 \frac{q_z^2}{q^2}  \sin \alpha \cos \alpha - \frac{q_y^2}{q^2}
 \cos \alpha   \sin \alpha + \frac{q_y  q_z}{q^2}  ( \cos^2 \alpha -
 \sin^2 \alpha ) - p_y  \cos \alpha - p_z  \sin \alpha \Biggr] .
\end{eqnarray}
In the same way we find for the relation of the spin correlation
functions in the reference frames defined with respect to the initial
muon beam polarization and the easy axis of magnetization,
respectively,
\begin{eqnarray}
\Phi^{\hat{x} \hat{x}}({\bf q}) & = & \Phi^{x x}({\bf q}) , \\
\Phi^{\hat{x} \hat{y}}({\bf q}) & = & \Phi^{\hat{y} \hat{x}}({\bf q})
 =  \Phi^{x y}({\bf q})  \cos \alpha
+  \Phi^{x z}({\bf q})  \sin \alpha , \\
\Phi^{\hat{x} \hat{z}}({\bf q}) & = & \Phi^{\hat{x} \hat{z}}({\bf q})
 =  \Phi^{x z}({\bf q})  \cos \alpha
 - \Phi^{x y}({\bf q})  \sin \alpha  , \\
\Phi^{\hat{y} \hat{y}}({\bf q}) & = & \Phi^{y y}({\bf q})  \cos^2 \alpha
+  \Phi^{z z}({\bf q})  \sin^2 \alpha  + 2 \Phi^{y z}({\bf q})
\cos \alpha  \sin \alpha , \\
\Phi^{\hat{y} \hat{z}}({\bf q}) & = & \Phi^{\hat{z} \hat{y}}({\bf q})
= \Phi^{z z}({\bf q})  \sin \alpha  \cos \alpha
- \Phi^{y y}({\bf q})  \sin \alpha  \cos \alpha
+ \Phi^{y z}({\bf q}) (\cos^2 \alpha - \sin^2 \alpha ) , \\
\Phi^{\hat{z} \hat{z}}({\bf q}) & = & \Phi^{y y}({\bf q})  \sin^2
\alpha +  \Phi^{z z}({\bf q})  \cos^2 \alpha  -
2 \Phi^{y z}({\bf q})
\cos \alpha  \sin \alpha .
\end{eqnarray}

\begin{multicols}{2}
\narrowtext

In section \ref{solution_mc} we have calculated the spin correlation
functions $\bar{\Phi}^{\alpha \alpha}({\bf q})$ in the eigenvector
basis ${\bf e}_\alpha ({\bf q})$. Upon using the transformation,
Eq.~\ref{transformation}, between cartesian spin components and the
spin components in the eigenvector basis we find
\begin{equation}
\Phi^{\beta \gamma}({\bf q}) = \sum_{\alpha} w_{\alpha \beta}({\bf q})
 w_{\alpha \gamma}({\bf q}) \bar{\Phi}^{\alpha \alpha}({\bf q}) .
\end{equation}
With the Fluctuation-Dissipation-Theorem the spin correlation functions
$\bar{\Phi}^{\alpha \beta}({\bf q})$ are given in terms of the statical
susceptibility and the linewidth,
\begin{eqnarray}
\bar{\Phi}^{\alpha \alpha} ({\bf q})& = & \frac{2 k_B
T}{\mu_0 (g_L \mu_B)^2} \frac{\chi_{\alpha}({\bf q})}
{\Gamma^{\alpha}({\bf q})} \nonumber \\
& = & \frac{2 k_BT}{J \mu_0 (g_L \mu_B)^2 A q^{z + 2}}
\frac{1}{\hat{\lambda}_{\alpha} ({\bf R}) \, \gamma_{\alpha}(\nu; {\bf R})} .
\end{eqnarray}
The muon relaxation rate $\lambda_z$ depends on the material parameters $q_{_A}
\xi_0$ and $q_{_D} \xi_0$ characterizing the strength of the uniaxial
anisotropy and the dipolar interaction, respectively. If we assume that the
uniaxial anisotropy in Gd is solely due to the combined effect of dipolar
interaction and non cubic lattice structure, one can estimate the ratio of
dipolar to uniaxial wave vector \cite{fujiki:87}
\begin{equation}
\frac{q_{_D}}{q_{_A}} = 7.8738 .
\end{equation}
With this assumption the number of material parameters is reduced to one,
$q_{_D} \xi_0$. In comparing our theory with $\mu$SR experiments at a
polarization $\alpha = 90^o$ we get the best fit to the data with $q_{_D} \xi_0
= 0.13$. This results in the following values for the uniaxial and dipolar wave
vector
\begin{eqnarray}
q_{_A}  =  0.0165 / \xi_0 , \quad
q_{_D}  = 0.13 / \xi_0 .
\end{eqnarray}
The corresponding crossover temperatures, $q_{A,D} \xi =1$, are given by
\begin{eqnarray}
T_A =  T_c + 0.43 \,  {\rm K}  , \quad
T_D =  T_c + 16.54 \, {\rm K}.
\end{eqnarray}
These set of parameters suggest the following crossover scenario. For $T \gg
T_D$ we expect critical behavior dominated by the (isotropic) Heisenberg fixed
point. The relaxation rate shows power law behavior
\begin{equation}
\lambda \propto t^{-w} ,
\end{equation}
with an exponent $w_{\rm I} \approx \nu (z-1) \approx 1$. For
temperatures in the interval $T_D > T > T_A$ dipolar interaction
becomes important. But, from our analysis of the uniaxial crossover in
section \ref{solution_mc} we have seen 
that the uniaxial crossovers in dynamics sets
in at wave vectors much larger than expected from an analysis of the
static quantities; i.e., the dynamic crossover in the longitudinal
scaling function right at $T_c$ is located at $q_{\rm cross} \approx
10 \times q_{_A}$. Therefore, even for $T>T_A$ we expect to observe
effects from dipolar interaction as well as uniaxial anisotropy.
Finally, for $T < T_A$ the critical dynamics is determined by the
uniaxial dipolar fixed point. Then the static susceptibilities do no
longer diverge for $q \rightarrow 0$ and $T \rightarrow T_c$ except
when the wave vector ${\bf q}$ is perpendicular to the easy axis of
magnetization. Since the relaxation rate $\lambda_z$ is given by an
integral over the whole Brillouin zone, the relative weight of the
critical axis along which the susceptibility diverges becomes
vanishingly small. As a consequence the relaxation rate $\lambda_z$ no
longer diverges for $T \rightarrow T_c$.

\begin{figure}[htb]
\centerline{\epsfig{figure=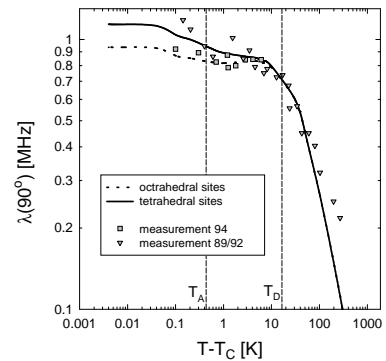,height=5.0cm,clip=}}
\caption{Experimental and theoretical results of the relaxation rate
  $\lambda$ for tetrahedral and octahedral muon sites with $\alpha = 90^o$.}
\label{figure16}
\end{figure}

\begin{figure}[htb]
\centerline{\epsfig{figure=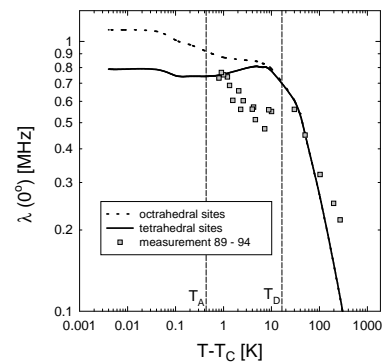,height=5.0cm,clip=}}
\caption{Experimental and theoretical results of the relaxation rate 
  $\lambda$ for tetrahedral and octahedral muon sites with $\alpha = 0^o$.}
\label{figure17}
\end{figure}

Since the interaction between the spin of the muon and the lattice
spins is a combination of Fermi contact and dipolar interaction, the
muon relaxation rate is a rather complicated function of the
relaxation rates along the eigendirections. Therefore, the temperature
dependence of the relaxation rate can no longer be described in terms
of simple power laws but shows a more complicated functional
dependence. In Figs.~\ref{figure16} and \ref{figure17} we show a
comparison between the theoretical and experimental results for two
different initial polarizations. 
In Fig.~\ref{figure16} the initial
polarization is inclined by an angle $\alpha = 90^o$ with respect to
the easy axis of magnetization. The solid line and dashed line are the
theoretical result for the muon relaxation rate if the muons
penetrating the sample are located at tetrahedral and octahedral
interstitial sites, respectively. The comparison between theory and
experiment favors tetrahedral sites. This is confirmed by $\mu$SR
experiments with the initial polarization along the easy axis of
magnetization (see Fig.~\ref{figure17}). The ratio $\lambda_z (90^o) /
\lambda_z (0^o)$ for $T \rightarrow T_c$ becomes $1.2$ and $0.7$ for
octahedral and tetrahedral sites, respectively. The experiment is
closer to the latter value.

\subsection{PAC and MS measurements}

Here we consider those hyperfine interaction probes (ME, PAC, NMR),
where the Fermi contact term gives the dominant contribution to the
hyperfine field at the nucleus. Then the corresponding interaction
Hamiltonian reduces to
\begin{equation}
H(t) = A_{\rm Fermi} \, {\bf I} \cdot {\bf S}(t)
\end{equation}
where ${\bf I}$ denotes the nuclear and ${\bf S}$ the electronic spin.
In hyperfine interaction experiments one observes the nuclear
relaxation rate due to the surrounding fluctuating electronic magnetic
moments. The standard experiments are performed in the motional
narrowing regime, where the nuclear relaxation rate $\tau_R^{-1}$ is
directly proportional to the (averaged) spin autocorrelation time
$\tau_c$
\begin{equation}
\tau_c = \frac{1}{2} \int_{- \infty}^{\infty} dt \frac{1}{3} \sum_{\alpha}
\Phi^{\alpha \alpha} ({\bf x} = 0,t),
\end{equation}
where the spin autocorrelation function is given by
\begin{equation}
\Phi^{\alpha \alpha}({\bf R},t) = \frac{1}{2} \Bigl \langle \{
S^{\alpha}({\bf x},t), S^{\alpha}({\bf 0},0) \} \Bigr \rangle.
\end{equation}
Hence the above hyperfine interaction methods provide an integral
property of the spin-spin correlation function. Upon using the
fluctuation dissipation theorem (FDT) we get
\begin{eqnarray}
\tau_c  = \frac{k_B \, T}{V_q} \int_{\rm BZ} d^3 {\bf q} \,
\frac{1}{3} \sum_{\alpha}
\frac{\chi_{\alpha}({\bf q};\xi,q_{_D},q_{_A})}
     {\Gamma_{\alpha}({\bf q};\xi,q_{_D},q_{_A})} .
\label{tau_c}
\end{eqnarray}
The ${\bf q}$-integration extends over the Brillouin zone (BZ), the
volume of which is $V_q$.

Important information about the behavior of the auto-correlation time
can be gained from a scaling analysis. Upon using the static and
dynamic scaling laws Eq.~(\ref{tau_c}) can be written as
\begin{equation}
  \tau_c \propto 4 \pi \int dq q^{-z} {1 \over 3} \sum_{\alpha}
  \frac{{\hat \chi}_{\alpha}
  \left({\hat {\bf q}}; q \xi,q/q_{_D},q/q_{_A} \right)}
       {\gamma_{\alpha}
  \left({\hat {\bf q}}; q \xi,q/q_{_D},q/q_{_A} \right)} \, ,
\label{tau_c_scal1}
\end{equation}
where we have neglected the Fisher exponent $\eta$. If there were no
dipolar interaction and no uniaxial anisotropy, one could extract the
temperature dependence from the integral in Eq.~(\ref{tau_c_scal1})
with the result $\tau_c \propto \xi^{z-1}$.  This expression can be
used to define an effective dynamical exponent $z_{\rm eff} (\tau)$,
which depends on the correlation length by
\begin{equation}
  \tau_c \propto \xi^{z_{\rm eff}-1} \propto
  \left( {T- T_c \over T_c} \right)^{-w_{\rm eff}} \, .
\label{tau_c_eff}
\end{equation}
with $w_{\rm eff} = \nu (z_{\rm eff} -1)$.

If dipolar interaction and uniaxial anisotropy are absent, one would
expect to observe a critical exponent for the relaxation rate $w
\approx 0.70 \times (5/2-1) \approx 1.0$. Dipolar interaction is known
to be a relevant perturbation with respect to the Heisenberg fixed
point; it leads to asymptotic static critical exponents which are only
slightly different from the corresponding Heisenberg values, but since
dipolar interaction implies a non-conserved order parameter the
asymptotic dynamic exponent becomes $z_{\rm D} \approx 2$.  Hence
dipolar interaction would induce a crossover from $w_{\rm I} \approx
1.0$ to $w_{\rm D} \approx 0.7$. Uniaxial interaction is also known to
be a relevant perturbation with respect to the Heisenberg fixed point.
Again, the static critical exponents are not changed very much, e.g.
one finds $\nu_{\rm I} = 0.63$, but the dynamic exponent becomes
$z_{\rm I} \approx 4$ if the order parameter is conserved ($z_{\rm I}
\approx 2$ otherwise). The corresponding exponent for the hyperfine
relaxation rate would turn out to be $w_{\rm I} \approx 1.89$ and
$w_{\rm I} \approx 0.63$ for conserved and non-conserved order
parameter, respectively. According to these scaling arguments it is
hard to think of any dynamic universality class which could lead to an
effective exponent $w_{\rm eff}$ smaller than about $0.6$. Actually,
however M\"ossbauer studies and PAC measurements on Gd show distinctly
anomalous low values $w \approx 0.5$, which can not be explained by
either of the above scenarios. This experimental puzzle can be
resolved if one considers the combined effect of dipolar interaction
and uniaxial anisotropy. As we have seen in our analysis of the static
critical behavior of uniaxial dipolar ferromagnets, {\em all} the
eigenvalues of the susceptibility matrix remain finite upon
approaching the critical temperature except when the wave vector of
the spin fluctuations is perpendicular to the easy axis of
magnetization. Since this is only a region of measure zero in the
Brillouin zone one actually expects that the relaxation rate
Eq.~(\ref{tau_c}) does no longer diverge upon approaching $T_c$, i.e.,
$w_{\rm UD} = 0$. For a quantitative comparison with the experiment
\cite{chowdhury-collins-hohenemser:84,collins-chowdhury-hohenemser:86,%
  chowdhury-collins-hohenemser:86} we rewrite the auto correlation
time in scaling form
\begin{eqnarray}
\tau_c
& = & \frac{2 \pi \, k_B \, T}{J \, V \, A}
\left(\xi^{-2} \! + \! q_{_D}^2 \! + \! q_{_A}^2 \right)^{-3/4}
\int_{R_0}^{\infty} d R \,  R^{1/2} \nonumber \\
&& \times \int_0^{\pi} d \nu \sin \nu \, \frac{1}{3} \sum_{\alpha}
\frac{1}{\hat{\lambda}_{\alpha}(\nu;{\bf R}) \, \gamma^{\alpha}(\nu;{\bf R})},
\label{tau_c_scal2}
\end{eqnarray}
where we have introduced polar coordinates. The lower cutoff is given by
\begin{equation}
R_0 = \frac{1}{q_{_{\rm BZ}}} \sqrt{\xi^{-2} + q_{_D}^2 + q_{_A}^2} ,
\end{equation}
where $q_{\rm BZ}$ is the boundary of the Brillouin zone. In the
critical region it can be disregarded and replaced by $R_0=0$, since
$q_{_{\rm BZ}} \gg q_{_D}, \, q_{_A}$ and the integrand in
Eq.~(\ref{tau_c_scal2}) is proportional to $\sqrt{R}$ for small $R$.
For very small $\xi$ (outside the critical region) the cutoff reduces
the autocorrelation time with respect to the critical value.  One
should note, that the dominant wave vectors contributing to the
relaxation time $\tau_c$ in Eq.~(\ref{tau_c_scal2}) are close to the
zone center \cite{reotier-yaouanc:94}.

\begin{figure}[htb]
\centerline{\epsfig{figure=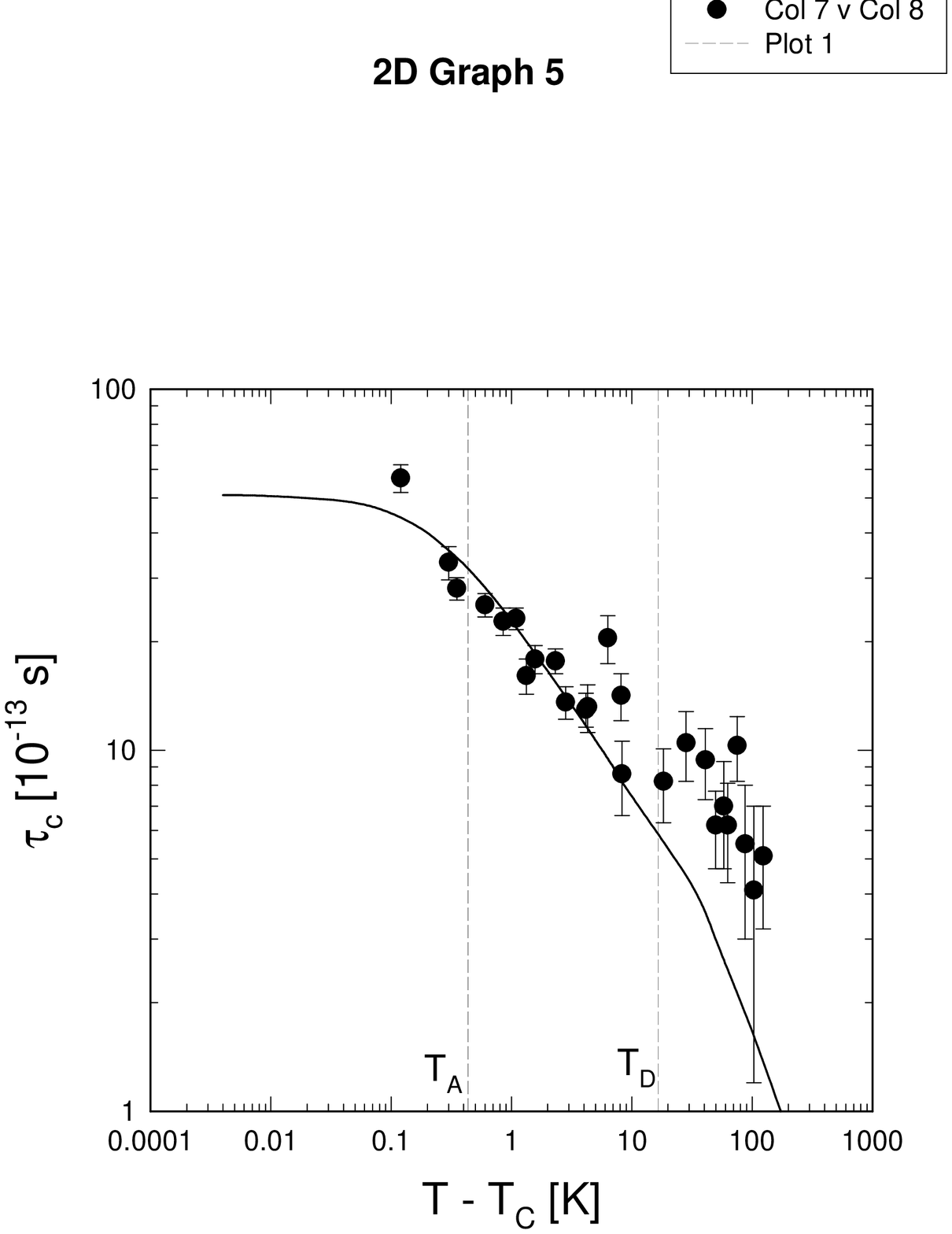,height=5.0cm,clip=}}
\caption{Experimental and theoretical results of the autocorrelation time
  $\tau_c$ for PAC experiments.}
\label{figure18}
\end{figure}

\begin{figure}[htb]
\centerline{\epsfig{figure=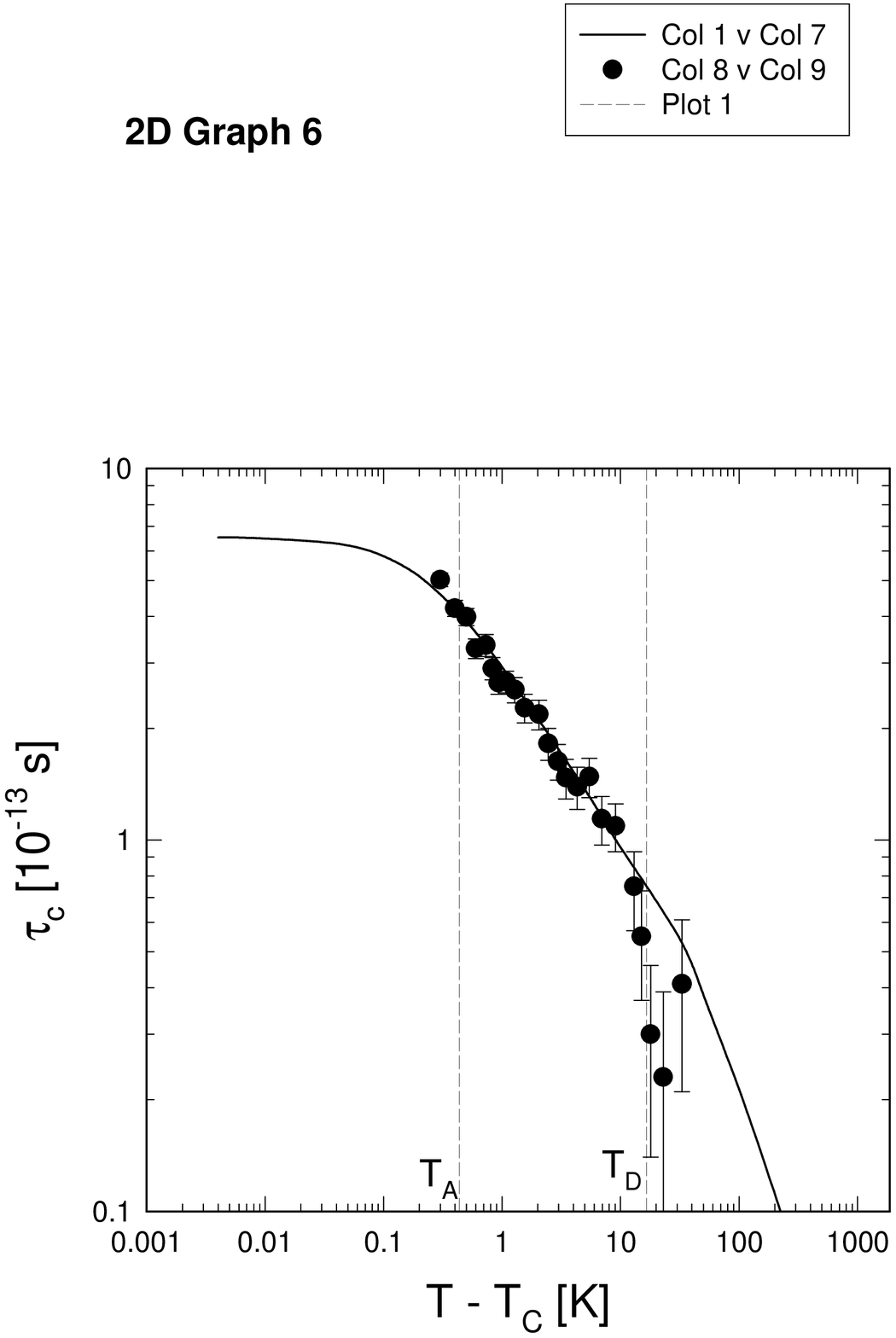,height=5.0cm,clip=}}
\caption{Experimental and theoretical results of the autocorrelation time
  $\tau_c$ for M{\"o}ssbauer spectroscopy experiments.}
\label{figure19}
\end{figure}

Let us now compare with hyperfine experiments on Gd mentioned above
\cite{collins-chowdhury-hohenemser:86,chowdhury-collins-hohenemser:84,%
  chowdhury-collins-hohenemser:86}.  The autocorrelation time $\tau_c$
is shown in Figs.~\ref{figure18} and \ref{figure19} for PAC
experiments and M\"ossbauer spectroscopy, respectively.  Both set of
data are in good agreement with the results from mode coupling theory
for $T-T_c < 10K$. Note that besides the overall frequency scale there
is no fit-parameter, since we have used the same set of values for the
dipolar and uniaxial wave vector as for our comparison with $\mu$SR
experiments. At higher temperatures the PAC data are above and the
M\"ossbauer data below the theoretical prediction.

The dipolar and uniaxial crossover temperatures indicated in the
figures show that all the data are within the crossover regime between
isotropic dipolar and uniaxial dipolar critical behavior. This
explains the anomalous low value for the effective exponent $w \approx
0.5$. This result clearly shows for the first time that the {\em
  universality class} for the critical behavior of Gd is the {\em
  uniaxial dipolar ferromagnet}.

\section{Summary and Conclusions}
\label{summary}

We have studied the critical dynamics of three-dimensional
ferromagnets with uniaxial anisotropy taking into account exchange and
dipole-dipole interaction. This analysis was mainly motivated by the
puzzling experimental situation for the rare earth material Gd.  Due
to the fact that it is an S-state ion one would expect that it should
have a vanishing magnetocrystalline anisotropy and therefore be an
almost ideal model system for an isotropic Heisenberg ferromagnet
(model J). But, the actual experimental observation shows large
deviations from the expected behavior. As explained in detail in the
main text, the way out of this puzzle is to realize that there is
always a dipolar interaction between the magnetic moments in the
material which in conjunction with the crystalline anisotropy can lead
to anisotropy in the magnetic properties of the system. 

The dipolar interaction and uniaxial anisotropy introduce two length
scales, the dipolar wave vector $q_{_D}$ and the wave vector $q_{_A}$
measuring the uniaxial anisotropy. Note that there are two sources for
the uniaxial anisotropy: magnetocrystalline (non S-state character of
the magnetic ions) and dipolar.  For Gd there is only the latter and
hence $q_{_D}$ and $q_{_A}$ are not really independent but their ratio is
fixed, $q_{_D}/q_{_A} = 7.87$. Both length scales then have a common source,
the dipole-dipole interaction. 

The presence of these length scales leads to a quite complicated
crossover scenario already for the static spin-spin correlation
function. The essence of the physics, however, can already be seen
within an Ornstein-Zernike approximation. Without uniaxial anisotropy
two out of the three eigenmodes of the correlation function are
critical, whereas the third correlation function starts to saturate
once the wave vector becomes smaller than the dipolar wave vector
$q_{_D}$. It is quite remarkable that due to the combined effect of
dipolar and uniaxial anisotropy {\em all} eigenvalues of the
susceptibilty matrix remain finite in the long wave length limit and
upon approaching the critical temperature. Only if the angle
$\nu$ between the easy axis of magnetization and the wave vector
is $\nu = 90^o$ the third eigenvalue (corresponding to the
transverse mode when uniaxiality is neglected) becomes critical.

We have used mode coupling theory to derive a set of integral
equations for the time-dependent spin-spin correlation functions.
Using the Lorentzian approximation we have solved these equations
numerically and determined the complete dynamic crossover scenario.
Spezializing to Gd with a relatively weak uniaxial anisotropy (induced
by the dipolar interaction) we find that the scaling functions for the
line widths show a behavior which is similar to what is found for cubic
dipolar systems (with no uniaxial anisotropy). The deviations become
largest as the critical temperature is approached.

In the final section we have compared our results with $\mu$SR, PAC
and M\"ossbauer experiments and find quantitative agreement with our
theory. This explains the anomously low values of the critical
exponents found in these experiments as a combined effect of lattice structure
and dipolar interaction.  From the quantitative agreement
between theory and experiment the following conclusions can be drawn:
\par (i) The universality class of Gd is the uniaxial dipolar
ferromagnet. \par (ii) The dominant factor for the uniaxial anisotropy
in Gd is solely the dipolar interaction. \par (iii) The theory even
allows us to predict that muons in Gd are located at tetrahedral
interstitial sites close to $T_c$. We expect that the analysis presented
in this paper will be valuable for interpreting dynamic measurements on
the non-cubic magnetic system.

\acknowledgements It is a pleasure to acknowledge helpful discussions
with A.  Yaouanc and P. Dalmas de R\'eotier. This work has been
supported by the German Federal Ministry for Education and Research
(BMBF) under contract number 03-SC5-TUM0 and be the Deutsche
Forschungsgemeinschaft under contract number SCHW~348/10. E.F.
acknowledges support from the Deutsche Forschungsgemeinschaft through
a Heisenberg fellowship (Fr.~850/3).

\appendix

\section{Zero-field $\mu$SR depolarization rate and spin correlation functions}
\label{appendix}

It can be shown \cite{henneberger:96} that the muon damping rate in
longitudinal geometry is related to the spin-spin correlation function
of the host material by a sum over all lattice sites
\begin{eqnarray}
\lambda_z
&=& \frac{\pi {\cal D}}{V^2}
    \sum_{i_1 i_2} \sum_{\beta \gamma}
    \Biggl[ G^{x \beta}({\bf r}_{i_1}) G^{x \gamma}({\bf r}_{i_2})
    \Phi_{i_1 i_2}^{\beta \gamma} \nonumber \\
&&  \quad + G^{y \beta}({\bf r}_{i_1}) G^{y \gamma}({\bf r}_{i_2})
    \Phi_{i_1 i_2}^{\beta \gamma} \Biggr]
\label{lambda_z_real_space}
\end{eqnarray}
where ${\bf r}_i$ is the vector pointing from the interstitial
position of the muon to the lattice site $i$. Here we have defined
${\cal D} = (\mu_0 / 4 \pi)^2 \gamma_\mu^2 (g_L \mu_B)^2$. Note that
the sum is weighted by the coupling tensor $G^{\alpha \beta}({\bf
  r}_i)$ of the muon and the spins of the magnet,
\begin{equation}
G^{\alpha \beta}({\bf r}_i)  =
D^{\alpha \beta}({\bf r}_i) + n_{\mu} \, H ({\bf r}_i) \delta_{\alpha \beta},
\end{equation}
with the dipolar contribution
\begin{eqnarray}
  D^{\alpha \beta}({\bf r}_i)
= v_a
  \left[
    \frac{3 r_i^\alpha r_i^\beta}{r_i^5} -
    \frac{\delta_{\alpha \beta}}{r_i^3}
  \right]
\end{eqnarray}
and the Fermi contact contribution $H_{{\bf r}_i}$, which equals $H$
for nearest-neighbor atoms and zero otherwise. The number of nearest
neighbors equals $n_\mu$. $\Phi_{ij}^{\alpha \beta}$ is the symmetrized
spin-spin correlation function between spins at sites $i_1$ and $i_2$
and zero frequency.
\begin{eqnarray}
\Phi_{i_1 i_2}^{\beta \gamma}
\!=\!  \frac{1}{2}
   [ \langle S_{i_1}^{\beta}(\omega=0)  S_{i_2}^{\gamma} \rangle \!+\!
     \langle S_{i_2}^{\gamma}(\omega=0) S_{i_1}^{\beta}(\omega=0) \rangle ]
\end{eqnarray}
Now we are going to rewrite Eq.~(\ref{lambda_z_real_space}) in Fourier space.
This leads to the evaluation of dipole sums
\begin{eqnarray}
  D^{\alpha \beta} ({\bf q)} =
  \sum_i D^{\alpha \beta} ({\bf r}_i) e^{i {\bf q} \cdot {\bf r}_i}.
\end{eqnarray}
How one deals with such dipole sums for cubic lattices has been
described in Ref.~\cite{yaouanc-reotier-frey:93}. Here we have to
calculate the dipole sum for lattices with a hcp structure. This is
most conveniently done by decomposing the lattice into four
orthorombic sublattices such that each site is in exactly one
sublattice \cite{fujiki:87}. The basis lattice vectors of the
sublattices in terms of the basis vectors of the hcp lattice are given
by
\begin{eqnarray}
{\bf a}_1 =  {\bf a} , \quad
{\bf a}_2 =  2 {\bf b} + {\bf a} , \quad \text{and} \quad
{\bf a}_3  =  {\bf c} .
\end{eqnarray}
The  volume of an orthorhombic unit cell $v_{\rm ortho}$ is four times the
volume $v_{hcp}$ per atom in gadolinium,
\begin{equation}
v_{\mathrm{hcp}} = \frac{1}{4} v_{\mathrm{ortho}}.
\end{equation}
A convenient choice of the origins of the four sublattices in terms of
the basis vectors ${\bf a}_1$, ${\bf a}_2$ and ${\bf a}_3$ is
\begin{eqnarray}
{\bf R}_0 & = & 0 \\
{\bf R}_1 & = & \frac{1}{2} {\bf a}_1 + \frac{1}{2} {\bf a}_2 \\
{\bf R}_2 & = & \frac{1}{3} {\bf a}_2 + \frac{1}{2} {\bf a}_3 \\
{\bf R}_3 & = & \frac{1}{2} {\bf a}_1 + \frac{5}{6} {\bf a}_2 +
                \frac{1}{2} {\bf a}_3.
\end{eqnarray}
The muon can either be located at a tetrahedral site
\begin{equation}
{\bf r}_0 = \frac{3}{8} {\bf a}_3 ,
\end{equation}
or a octahedral site
\begin{equation}
{\bf r}_0 = \frac{1}{2} {\bf a}_1 + \frac{1}{6} {\bf a}_2 +
\frac{1}{4} {\bf a}_3 .
\end{equation}

\end{multicols}
\widetext

Putting things togehter we find
\begin{eqnarray}
\lambda_z  = \frac{\pi {\cal D}}{V^2}
\sum_{l_1  l_2} \sum_{i_1 i_2} \sum_{\beta \gamma}
\left[
  G^{x \beta}  ({\bf r}_0 \! + \!{\bf i}_1 \! +  \!{\bf R}_{l_1})
  G^{x \gamma} ({\bf r}_0 \! + \!{\bf i}_2 \! +  \!{\bf R}_{l_2})
+ G^{y \beta}  ({\bf r}_0 \! + \!{\bf i}_1 \! +  \!{\bf R}_{l_1})
  G^{y \gamma} ({\bf r}_0 \! + \!{\bf i}_2 \! +  \!{\bf R}_{l_2})
\right]
\Phi^{\beta \gamma}_{{\bf i}_1 + {\bf R}_{l_1}, {\bf i}_2 + {\bf R}_{l_2}}
\end{eqnarray}
Introducing Fourier transformed spin variables for each of the sublattices
\begin{equation}
S^{\alpha}({\bf r}_0 + {\bf i} + {\bf R}_{l})
= \frac{1}{N} \sum_{{\bf q}}
  \exp[i {\bf q} \cdot ({\bf r}_0 + {\bf i} + {\bf R}_l) ] \,
  S^{\alpha}_l ({\bf q})
\end{equation}
where ${\bf R}_l$ are the lattice vectors of the $l$-th sublattice, one gets
\begin{eqnarray}
\lambda_z  =  \pi \gamma_{\mu}^2 \left( \frac{\mu_0}{4 \pi} \right)^2
\frac{(g_L \mu_B)^2}{V^2}
\sum_{{\bf q}} \sum_{\beta \gamma} \left[ G^{x \beta}({\bf q})
G^{x \gamma}(-{\bf q}) + G^{y \beta}({\bf q}) G^{y \gamma}(-{\bf q})
\right] \Phi^{\beta \gamma}({\bf q}) ,
\end{eqnarray}
where the Fourier transform for the symmetrized spin-spin correlation function
and the coupling tensor reads
\begin{eqnarray}
G^{\alpha \beta}({\bf q}) & = & \sum_{l=0}^3 \sum_i
G^{\alpha \beta}({\bf r}_0 + {\bf i} + {\bf R}_{l})
\exp[i {\bf q} \cdot ({\bf r}_0 + {\bf i} + {\bf R}_{l})] \\
\Phi^{\beta \gamma}_{{\bf i}_1 + {\bf R}_{l_1},
{\bf i}_2 + {\bf R}_{l_2}} & = & \frac{1}{N^2} \sum_{{\bf q}}
\exp[i {\bf q} \cdot ({\bf i}_1 - {\bf i}_2)]
\exp[i {\bf q} \cdot ({\bf R}_{l_1} - {\bf R}_{l_2})]
\Phi^{\beta \gamma } ({\bf q}) \\
& = & \frac{1}{N^2} \sum_{{\bf q}}
\exp[i {\bf q} \cdot (({\bf r}_0 + {\bf R}_{l_1} + {\bf i}_1)
 - ({\bf r}_0 + {\bf R}_{l_2} + {\bf i}_2))]
\Phi^{\beta \gamma } ({\bf q}) \\
\Phi^{\beta \gamma } ({\bf q})
& = & \frac{1}{2}
\left[
\langle S^{\beta}_{\bf q} (\omega = 0)
        S^{\gamma}_{- {\bf q}} (\omega = 0)
\rangle +
\langle S^{\gamma}_{-{\bf q}} (\omega = 0)
        S^{\beta}_{\bf q} (\omega = 0)
\rangle
\right]
\end{eqnarray}

\begin{multicols}{2}
\narrowtext

The summation over ${\bf q}$ runs over the $N$ vectors in the first Brillouin
zone. For sufficiently large $N$ the summation may be replaced by an integral
\begin{equation}
\lambda_z = \frac{ \pi {\cal D}} {V}
\int_{{\bf q}} \sum_{\beta \gamma}
\left[
G^{x \beta}_{\bf q} G^{x \gamma}_{-{\bf q}} +
G^{y \beta}_{\bf q} G^{y \gamma}_{-{\bf q}}
\right]
\Phi^{\beta \gamma}({\bf q}) .
\end{equation}
In principle the damping rate $\lambda_z$ depends on the wave vector
dependence of the coupling tensor and the symmetrized spin-spin
correlation function over the whole Brillouin zone. It was shown,
however, in Ref.~\cite{reotier-yaouanc-frey:94} that in $\mu$SR
measurements close to $T_c$ the dominant contribution to $\lambda$
comes from the vicinity of the Brillouin zone center. The coupling
tensor $G^{\alpha \beta}({\bf q})$ can therefore be approximated by
its limiting behavior near $q \to 0$.

Now we turn to the evaluation of the dipolar sums in the coupling
tensor.  The dipolar sums can be evaluated in Fourier space by the
method of Ewald summation. The sum in Fourier space is divided in a
sum over the direct lattice and a sum over the indirect lattice, so
that both sums converge quickly.

\end{multicols}
\widetext

With the decomposition into four orthorombic sublattices we get
\begin{eqnarray}
D^{\alpha \beta}({\bf q})& = & \sum_{l=0}^3 \sum_i D^{\alpha \beta}
({\bf r}_{i,l}) \, \exp\left[ i {\bf q} \cdot
({\bf r}_{i,l}) \right] \nonumber \\
& = & \sum_{l=0}^3 \sum_i D^{\alpha \beta}
({\bf i} + {\bf R}_l - {\bf r}_0) \, \exp\left[ i {\bf q} \cdot
({\bf i} + {\bf R}_l - {\bf r}_0) \right] \nonumber \\
& = & v_{{\rm hcp}} \sum_{l=0}^3
\exp[i {\bf q} \cdot ({\bf r}_0 + {\bf R}_l)] \left[
\frac{\partial^2}{\partial x_{\alpha} \, \partial x_{\beta}}
\left( \sum_i \frac{ \exp( i
{\bf q} \cdot {\bf i})}{|{\bf i} - {\bf x}|} \right)
\right]_{{\bf x} = {\bf r}_0 - {\bf R}_l}.
\end{eqnarray}
The latter expression can be evaluated in the same way as for cubic lattices
\cite{yaouanc-reotier-frey:93}. One finds \cite{henneberger:96}
\begin{eqnarray}
D^{\alpha \beta }({\bf q})& = & \frac{1}{4} \sum_{l=0}^3
D_l^{\alpha \beta }({\bf q})
=  - 4 \pi \left[\frac{q_{\alpha} q_{\beta}}{q^2} - \frac{1}{4}
\sum_{l=0}^3 C_l^{\alpha \beta}({\bf q}) \right]
=  - 4 \pi \left[\frac{q_{\alpha} q_{\beta}}{q^2} -
C^{\alpha \beta}({\bf q}) \right] ,
\end{eqnarray}
where
\begin{eqnarray}
C_l^{\alpha \beta}({\bf q}) & = & 4 \pi
\frac{q_{\alpha} q_{\beta}}{q^2} \left[ 1 - \exp \left( -\frac{q^2}{4 \rho^2}
\right) \right]
- \frac{1}{4 \rho^2} \sum_{{\bf K} \not = 0} (K_{\alpha} + q_{\alpha})
(K_{\beta} + q_{\beta}) \varphi_0 \left(\frac{({\bf q} + {\bf K} )^2}
{4 \rho^2} \right) \exp [- i {\bf K} \cdot ({\bf R}_l - {\bf r}_0)]
\nonumber \\
& & + \frac{v \rho^3}{2 \pi^{3/2}} \sum_i \Bigl[ 2 \rho^2
({\bf R}_l + {\bf i} - {\bf r}_0)_{\alpha}
({\bf R}_l + {\bf i} - {\bf r}_0)_{\beta} \, \varphi_{3/2}(\rho^2 r_i^2)
- \delta_{\alpha \beta} \, \varphi_{1/2}(\rho^2 r_i^2)\, \exp [i {\bf q} \cdot
({\bf R}_l + {\bf i}) - {\bf r}_0] \Bigr] .
\end{eqnarray}

\begin{multicols}{2}
\narrowtext

To lowest order in $q$ this reduces to
\begin{equation}
D^{\alpha \beta}( {\bf q} \to 0) =
- 4 \pi \left[ \frac{q_{\alpha} q_{\beta}}{q^2} -
C^{\alpha \beta} ({\bf q} = 0) \right]  ,
\end{equation}
with
\begin{equation}
C^{\alpha \beta}({\bf q} = 0) = \left( \begin{array}{ccc}
0.3485 & 0 & 0 \\
0 & 0.3485 & 0  \\
0 & 0 & 0.3030 \\
\end{array} \right)
\end{equation}
for octahedral and
\begin{equation}
C^{\alpha \beta}({\bf q} = 0) = \left( \begin{array}{ccc}
0.3118 & 0 & 0 \\
0 & 0.3118 & 0  \\
0 & 0 & 0.3764 \\
\end{array} \right)
\end{equation}
for tetrahedral sites.

\newpage

\begin{table}
\narrowtext
\begin{tabular}{ccccc}
 & $D^{xx}(=D^{yy})$ & $D^{zz}$ & $D^{xy}$ & $D^{xz}(=D^{yz})$ \\
\hline
$\beta_1^{\alpha \beta}$ &       &       & 0.44 & 0.53 \\
$\beta_2^{\alpha}$       & -0.16 & -0.11 &      &      \\
$\beta_1^{\alpha \alpha} - \beta_3^\alpha$
                         & 0.44  & 0.31  &      &      \\
$\beta_4^\alpha$         & 4.12  & 4.32  &      &      \\
$\beta_z$                & -0.06 &       &      &
\end{tabular}
\caption{Coefficients in the expansion of the dipolar tensor for a hcp
lattice structure with $c=5.78 \AA$, $a = 3.62 \AA$, and $c/a = 1.59$, as
appropriate for Gd bear $T_c$. Values are obtained from Ref.~\protect\cite{fujiki:87}
upon multipying with the volume of the primitive cell $v_a$.}
\label{coefficients_hcp}
\end{table}

\begin{table}
\narrowtext
\begin{tabular}{ccc}
region & $\Gamma_{\perp}$ & $\Gamma_{\parallel}$ \\
\hline
UC & $q_{_A}^{{5/2}}   $ & $q_{_A}^{-{3/2}} q^4 $  \\
UH & $q_{_A}^{{5/2}}  $ & $q_{_A}^{-{3/2}} \, \xi^{-2} \, q^2 $ \\
IC & $q^{{5/2}}$ & $q^{{5/2}} $ \\
IH & $\xi^{-{1/2}} q^2 $ & $\xi^{-{1/2}} q^2 $
\end{tabular}

\begin{tabular}{clcl}
C : & critical  & U : & uniaxial  \\
 H : & hydrodynamic  \hspace{0.5cm} & I : & isotropic \\
\end{tabular}

\caption{Analytic results for the asymptotic behavior of the transverse and
longitudinal line widths for aniotropic ferromagnets with a uniaxial
magneto-crystalline anisotropy, where the spins are coupled by short range
exchange interaction only.}
\label{table_asymptotic_uniaxial}
\end{table}

\begin{table}
\narrowtext
\begin{tabular}{ccc}
region & $\Gamma_{T}$ & $\Gamma_{L}$ \\
\hline
DC & $q_{_D}^{{1/2}}q^2$ & $q_{_D}^{{5/2}}  $  \\
DH & $q_{_D}^{{1/2}} \, \xi^{-2} \, $ & $q_{_D}^{{5/2}}  $ \\
IC & $q^{{5/2}}$ & $q^{{5/2}} $ \\
IH & $\xi^{-{1/2}} q^2 $ & $\xi^{-{1/2}} q^2 $
\end{tabular}

\begin{tabular}{clcl}
C : & critical  & D : & dipolar  \\
H : & hydrodynamic  \hspace{0.5cm} & I : & isotropic \\
\end{tabular}

\caption{Analytic results for the asymptotic behavior of the transverse and
longitudinal linewidths for isotropic dipolar ferromagnets, where the spins
are coupled by short-range exchange interaction and long-range dipolar
interaction.}
\label{table_asymptotic_dipolar}
\end{table}

\begin{table}
\narrowtext
\begin{tabular}{cccc}
region & $\Gamma_{1}$ & $\Gamma_{2}$ & $\Gamma_{3}$ \\
\hline
UDC  $\nu \not = 90^o$ & $ q_{_A}^{{5/2}} $
                       & $ q_{_A}^{{5/2}} $
                       & $ q_{_A}^{{5/2}} $ \\
UDC  $\nu = 90^o$      & $ q_{_A}^{{5/2}} $
                       & $ q_{_A}^{{5/2}} $
                       & $ q_{_A}^{{1/2}} \, q^2 $ \\
DUC  $\nu \not = 90^o$ & $ q_{_D}^{{1/2}} \, q_{_A}^2$
                       & $ q_{_D}^{{5/2}} $
                       & $ q_{_D}^{{5/2}} $ \\
DUC  $\nu = 90^o$      & $ q_{_D}^{{1/2}} \, q_{_A}^2 $
                       & $ q_{_D}^{{5/2}} $
                       & $ q_{_D}^{{1/2}} \, q^2 $ \\
UDH  $\nu \not = 90^o$ & $ q_{_A}^{{5/2}} $
                       & $ q_{_A}^{{5/2}} $
                       & $ q_{_A}^{{5/2}} $ \\
UDH  $\nu = 90^o$      & $ q_{_A}^{{5/2}} $
                       & $ q_{_A}^{{5/2}} $
                       & $ q_{_A}^{{1/2}} \, \xi^{-2} $ \\
DUH  $\nu \not = 90^o$ & $ q_{_D}^{{1/2}} \, q_{_A}^2  $
                       & $ q_{_D}^{{5/2}} $
                       & $ q_{_D}^{{5/2}} $ \\
DUH  $\nu = 90^o$      & $ q_{_D}^{{1/2}} \, q_{_A}^2 $
                       & $ q_{_D}^{{5/2}} $
                       & $ q_{_D}^{{1/2}} \, \xi^{-2} $ \\
UC                     & $ q_{_A}^{{5/2}} $
                       & $ q_{_A}^{{5/2}} $
                       & $ q_{_A}^{-{3/2}} \, q^4 $ \\
UH                     & $ q_{_A}^{{5/2}} $
                       & $ q_{_A}^{{5/2}} $
                       & $ q_{_A}^{-{3/2}} \, \xi^{-2} q^2 $ \\
DC                     & $ q_{_D}^{{1/2}}  q^2 $
                       & $ q_{_D}^{{5/2}} $
                       & $ q_{_D}^{{1/2}} q^2 $ \\
DH                     & $q_{_D}^{{1/2}} \, \xi^{-2} $
                       & $ q_{_D}^{{5/2}} $
                       & $ q_{_D}^{{1/2}} \,  \xi^{-2} $ \\
IC                     & $ q^{{5/2}} $
                       & $ q^{{5/2}} $
                       & $ q^{{5/2}} $ \\
IH                     & $ \xi^{-{1/2}} q^2 $
                       & $ \xi^{-{1/2}} q^2 $
                       & $ \xi^{-{1/2}} q^2 $
\end{tabular}

\begin{tabular}{clclcl}
C : & critical & D : & dipolar & H : &  hydrodynamic  \\
U : & uniaxial & I : & isotropic & & \\
\end{tabular}

\caption{Analytic results for the asymptotic behavior of the linewidths in the
eigendirections for ferromagnets with a uniaxial magneto-crystalline
anisotropy, where the spins are coupled by short-range exchange interaction
and long-range dipolar interaction.}
\label{table_asymptotic_general}
\end{table}

\end{multicols}

\end{document}